%% file: main.tex
\pdfoutput=1
\documentclass{article}
\usepackage{amsfonts,amsmath,amssymb,authblk,floatrow,fullpage,graphicx,tabularx,threeparttable,tikz,pgfplots,xcolor,lscape,multirow,verbatim,gensymb}
\usepackage[colorlinks]{hyperref}
\usepackage[noabbrev,capitalise,sort&compress]{cleveref}
\usepackage{subcaption}
\usepackage{appendix}
\usepackage[textsize = small]{todonotes}
\usepackage[nohyperlinks,nolist]{acronym}
\usepackage{import}
\usetikzlibrary{arrows,arrows.meta,calc,fit,spy}
\newfloatcommand{capbtabbox}{table}[][\FBwidth]
\input{commands.tex}

\pgfplotsset{compat=newest}
\begin{document}
\title{An efficient approach to include transport effects in thin coating layers in electrochemo-mechanical models for all-solid-state batteries}
\author{Stephan Sinzig$^{1,2}$, Christoph P. Schmidt$^{1}$, Wolfgang A. Wall$^{1}$}
\affil{\small $^1$ Technical University of Munich, Germany, TUM School of Engineering and Design, Institute for Computational Mechanics, Boltzmannstra\ss e 15, 85748 Garching bei M\"unchen \\ $^2$ TUMint.Energy Research GmbH, Lichtenbergstra\ss e 4, 85748 Garching bei M\"unchen, Germany}
\date{}
\maketitle
\section*{Abstract}
\label{sec:abstract}
\input{chapters/abstract.tex}
\section{Introduction}
\label{sec:introduction}
\input{chapters/chapter_1.tex}
\section{ASSB model including transport in thin coating layers}
\label{sec:problem_statement}
\input{chapters/chapter_2.tex}
\section{Numerical aspects of the model}
\label{sec:numerics}
\input{chapters/chapter_3.tex}
\section{Numerical examples}
\label{sec:results}
\input{chapters/chapter_4.tex}
\section{Summary}
\input{chapters/chapter_5.tex}
\section*{Funding}
\input{chapters/funding.tex}
\appendixtitleon
\setcounter{table}{0}
\renewcommand{\thetable}{\Alph{section}.\arabic{table}}
\setcounter{equation}{0}
\renewcommand{\theequation}{\Alph{section}.\arabic{equation}}
\setcounter{figure}{0}
\renewcommand{\thefigure}{\Alph{section}.\arabic{figure}}
\begin{appendices}
\input{chapters/appendix.tex}
\end{appendices}
%
%************************************************************************%
% bibliography
%************************************************************************%
\bibliographystyle{IEEEtran}
\bibliography{literature}
\end{document}

%% file: commands.tex
\renewcommand{\vec}[1]{\boldsymbol{#1}}         % for vectors
\newcommand{\mat}[1]{\boldsymbol{#1}}           % for matrices

\newcommand{\udefnp}{{\vec{u}}}
\newcommand{\lambdadefnp}{{\vec{\lambda}_{\vec{u}}}}
\newcommand{\cbulknp}{{\vec{c}_\text{bulk}}}
\newcommand{\potbulknp}{{\vec{\Phi}_\text{bulk}}}
\newcommand{\csurfnp}{{\vec{c}_\text{manif}}}
\newcommand{\potsurfnp}{{\vec{\Phi}_\text{manif}}}
\newcommand{\lambdacsurfnp}{{\vec{\lambda}_{\vec{c}}}}
\newcommand{\lambdapotsurfnp}{{\vec{\lambda}_{\vec{\Phi}}}}

\newcommand{\Rudefnp}{\mat{R}_{\vec{u}}}
\newcommand{\Rlambdadefnp}{{\mat{R}_{\vec{\lambda}_{\vec{u}}}}}
\newcommand{\Rcbulknp}{{\mat{R}_{\vec{c}}^\text{bulk}}}
\newcommand{\Rpotbulknp}{{\mat{R}_{\vec{\Phi}}^\text{bulk}}}
\newcommand{\Rcsurfnp}{{\mat{R}_{\vec{c}}^\text{manif}}}
\newcommand{\Rpotsurfnp}{{\mat{R}_{\vec{\Phi}}^\text{manif}}}
\newcommand{\Rlambdacsurfnp}{{\mat{R}_{\vec{\lambda}_{\vec{c}}}}}
\newcommand{\Rlambdapotsurfnp}{{\mat{R}_{\vec{\lambda}_{\vec{\Phi}}}}}
\newcommand{\Rcbulksurfnp}{{\mat{R}_{\vec{c}}^\text{bulk-manif}}}
\newcommand{\Rpotbulksurfnp}{{\mat{R}_{\vec{\Phi}}^\text{bulk-manif}}}

\definecolor{ccao}{rgb}{0.596, 0.251, 0.043}
\definecolor{ccaa}{rgb}{0.314, 0.882, 0.957}
\definecolor{ael}{rgb}{0.063, 0.714, 0.188}
\definecolor{elel}{rgb}{0.478, 0.953. 0.373}
\definecolor{elcoat}{rgb}{0.584, 0.106, 0.478}
\definecolor{coatc}{rgb}{0.427, 1.0, 0.420}
\definecolor{cccc}{rgb}{0.796, 0.145, 0.129}
\definecolor{elccc}{rgb}{0.216, 0.114, 0.781}
\definecolor{ccco}{rgb}{0.533, 0.322, 0.055}

\newcommand{\parder}[2]{\frac{\partial #1}{\partial #2}}

%% file: chapters/abstract.tex
A novel approach is presented to efficiently include transport effects in thin active material coating layers of all-solid-state batteries using a dimensionally reduced formulation embedded into a three-dimensionally resolved coupled electrochemo-mechanical continuum model. In the literature, the effect of coating layers is so far captured by additional zero-dimensional resistances to circumvent the need for an extremely fine mesh resolution. However, a zero-dimensional resistance cannot capture transport phenomena along the coating layer, which can become significant, as we will show in this work. Thus, we propose a model which resolves the thin coating layer in a two-dimensional manifold based on model assumptions in the direction of the thickness. This two-dimensional formulation is monolithically coupled with a three-dimensional model representing the other components of a battery cell. The approach is validated by showing conservation properties and convergence and by comparing the results with those computed with a fully resolved model. Results for realistic microstructures of a battery cell, including coating layers as well as design recommendations for a preferred coating layer, are presented. Based on those results, we show that existing modeling approaches feature remarkable errors when transport along the coating layer is significant, whereas the novel approach resolves this.

%% file: chapters/chapter_1.tex
All-solid-state batteries (ASSBs) are seen as a promising technology to overcome the physical limits conventional lithium-ion batteries face today~\cite{Janek2016}. Some of the expectations are higher power densities due to the negligible polarization of the solid electrolyte, higher energy densities by enabling the use of the lithium metal anode, and higher safety standards due to the non-flammability of many solid electrolytes compared to conventional lithium-ion batteries~\cite{Takada2013, Fu2022, Famprikis2019}. However, ASSBs are mostly still only at the lab scale and require more research effort for their usage in relevant applications (e.g. \cite{Janek2023}). \\
Especially, phenomena at the internal interfaces of an ASSB cell are crucial for the overall cell performance. The properties of the interface between the solid electrolyte and the electrode can be tuned by adding a thin coating layer between both domains~\cite{Karayaylali2019}. The transport properties inside this coating layer can significantly differ from those in the other domains and feature transport also along the coating layer. By selecting a material of the coating layer together with the material of the bulk electrolyte, an optimal hybrid electrolyte can be designed by making use of the advantageous properties of the two materials while overcoming their individual drawbacks: for example benefiting from the good ionic conductivity of thiophosphates, from the oxidation stability of oxides, and from the mechanical flexibility of polymers. In the literature, different combinations of active materials, coatings, and solid electrolytes are reported~\cite{Culver2019,Wen2017,Nisar2021,Kaur2022} and tested: coatings that are argyrodite-based~\cite{Ma2022}, polymer-based~\cite{Huang2021a}, LiPON-based~\cite{Shrestha2021}, or $\text{LiNbO}_3$-based~\cite{Walther2021}. However, realizing all combinations of the various materials requires considerable effort. Furthermore, the quantification of the influence of the coating layer on the cell performance by measurements is complex and requires the combination of various experimental methods~\cite{Moryson2021}. Here, simulation models provide unprecedented insight, can find optimal combinations of coating and bulk materials, and can systematically separate the physical phenomena inside coating layers. All this can even be done before coating materials are synthesized or established in a production process.\\
Different well-established simulation approaches for the analysis of coating layers exist in the literature. A prominent class of simulation approaches is the computation of coating layers with models using the density functional theory (DFT). These models are based on first principles and, thus, enable to determine e.g. the transport properties of materials. One drawback for real applications is that they are restricted to small domains (e.g. \cite{Hao2013, Xu2015}). Therefore, physics-based continuum models are seen as appropriate when investigating the behavior of an entire battery cell. \\
Different approaches to incorporate a thin coating layer into a continuum simulation model are conceivable, e.g. by resolving it spatially or by representing it by a zero-dimensional resistance~\cite{Javed2020, Neumann2020}. Both approaches have advantages but also drawbacks: A spatial resolution requires an extremely fine mesh due to the small thickness of the coating layer compared to its tangential dimensions and, thus, a huge computational effort. A zero-dimensional resistance cannot capture transport along the coating layer which could become relevant as soon as the conductivity of the coating material reaches the order of the conductivity of the bulk electrolyte, as we will show in this work. \\
Thus, we propose an efficient approach to resolve the thin coating layer on a two-dimensional manifold embedded into a three-dimensional geometry of a microstructure of a battery cell. The evolution of quantities in the thickness direction of the coating layer is modeled by a priori knowledge and is included by a linear model and does, therefore, not require to be discretized in space. Thereby, we overcome the drawbacks of other modeling approaches, as our model does not require a prohibitive fine discretization of both the coating layer and the adjacent domains but simultaneously allows for capturing the potentially relevant transport of charge and mass along the coating layer. A rough estimate of the computational effort (see Appendix) for a geometrically realistic microstructure with the novel approach and a fully resolved approach reveals a gain of two orders of magnitude in unknowns and hence, shows the relevance of avoiding simulations with fully resolved coating layers. Especially if advanced interface coupling algorithms for non-matching mesh discretizations, like mortar methods (see e.g. \cite{Fang2018}) are not available, also the domains attached to the coating layer require a fine discretization and the reduction of the computational effort becomes even more important. \\
This work is organized as follows: First, a coupled electrochemo-mechanical continuum model together with a novel model to capture transport in the coating layer in ASSBs is introduced. Afterwards, its numerical treatment is summarized, and numerical examples are presented.

%% file: chapters/chapter_2.tex
The performance of a ASSB cell is strongly influenced by the interaction between electrochemistry and solid mechanics. Thus, a simulation model for ASSBs needs to consider this interaction (see e.g. \cite{Schmidt2022}) in the different components of the battery cell. In this work, we focus on the transport of mass and charge in coating layers on active material particles. One geometric characteristic of the coating layer is that its thickness is significantly smaller than its tangential dimensions. The evolution of the transported quantity in the thickness direction is often a priori known or can very well be approximated. We use this knowledge to derive a computationally efficient formulation representing transport phenomena inside thin coating layers. The formulation is embedded into a three-dimensionally resolved coupled electrochemo-mechanical continuum model for ASSBs to allow for the investigation of the influence of the coating layer on the entire cell performance. We begin with summarizing the governing equations for the bulk domain of ASSBs. Afterwards, a novel approach is introduced to capture transport effects in coating layers. Finally, interface and boundary conditions are summarized. The symbols listed in \cref{table:list_of_symbols} are used.
\subsection{Geometric definitions}
\label{sec:geometric_definitions}
\begin{figure}[ht]
  \centering
  \def\svgwidth{\textwidth}
  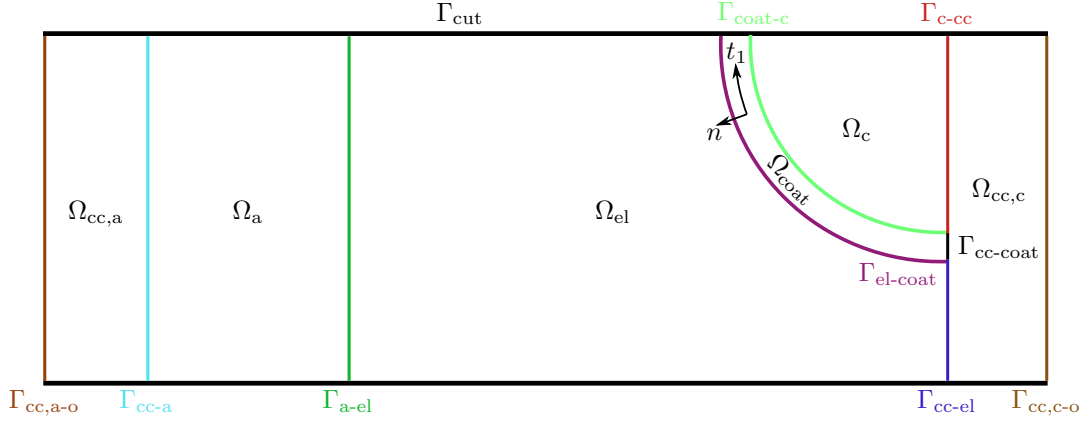
  \caption{Schematic sketch of the computational domain. The domain is split into subdomains $\Omega_i$. The surfaces~$\Gamma_{i-j}$ denote the interfaces between the domains~$\Omega_i$ and~$\Omega_j$.}
  \label{fig:domainsInterfaces}
\end{figure}

The domain of the battery cell is split into subdomains which geometrically define the components of a battery (see \cref{fig:domainsInterfaces}): The electrodes $\Omega_\text{ed} = \Omega_\text{c} \cup \Omega_\text{a}$ with cathode~$\Omega_\text{c}$ and anode~$\Omega_\text{a}$, the solid electrolyte~$\Omega_\text{el}$, the current collectors ~$\Omega_\text{cc} = \Omega_\text{cc,c} \cup \Omega_\text{cc,a}$ on the cathode side~$\Omega_\text{cc,c}$ and on the anode side~$\Omega_\text{cc,a}$, and the coating layer~$\Omega_\text{coat}$. Wherever two domains~$\Omega_i$ and~$\Omega_j$ intersect, an interface~$\Gamma_{i-j}$ is defined. The boundary is split into physically meaningful boundaries on the current collectors~$\Gamma_\text{cc-o} = \Gamma_\text{cc,c-o} \cup \Gamma_\text{cc,a-o}$ and boundaries where the geometry is cut to obtain a statistically representative domain~$\Gamma_\text{cut}$. A natural coordinate system is introduced in the coating domain with one coordinate~$n$ normal to the interface $\Gamma_\text{coat-c}$ and two coordinates~$t_{1,2}$ tangential to the interface $\Gamma_\text{coat-c}$.
\subsection{Governing equations for the bulk domain in a continuum formulation}
The model needs to satisfy the conservation of mass, charge, and linear momentum inside the battery cell, as discussed in literature~\cite{Schmidt2022}. This work only summarizes the corresponding equations and the coupling constraints between the fields for brevity (see Appendix for the derivation). The governing equations for the solid mechanics field are
\begin{subequations}
\begin{align}
    \nabla_\mathbf{X} \cdot (\mat{F} \mat{S}) + \vec{b}_0 &= \rho_0 \ddot{\vec{u}} \quad \text{in} \ \Omega_0, \\
    \mat{F} &= \mat{F}_\text{el} \mat{F}_{\text{growth}}, \\
    \mat{S} &= 2 \ \text{det}(\mat{F}_\text{growth}) \mat{F}_\text{growth}^{-1} \parder{\Psi_\text{el}}{\mat{C}_\text{el}} \mat{F}_\text{growth}^{-\text{T}}.
\end{align}
\end{subequations}
The governing equations for the electrochemical field are
\begin{subequations}
\begin{alignat}{2}
    \nabla \cdot (- \sigma \nabla \Phi) &= 0 &&\quad \text{in} \ \Omega_\text{ed} \cup \Omega_\text{cc},
    \label{eq:pot_solid_electrode} \\
    \nabla \cdot (- \kappa \nabla \Phi) &= 0 &&\quad \text{in} \ \Omega_\text{el} \cup \Omega_\text{coat},
    \label{eq:pot_solid_electrolyte} \\
    \left.\parder{c}{t}\right|_{\vec{X}} + c \ \nabla \cdot \dot{\vec{u}} - \nabla \cdot (D \nabla c) &= 0 &&\quad \text{in} \ \Omega_\text{ed}, \\
    \left.\parder{c}{t}\right|_{\vec{X}} + c \ \nabla \cdot \dot{\vec{u}} &= 0 &&\quad \text{in} \ \Omega_\text{el} \cup \Omega_\text{coat}.
\end{alignat}
\end{subequations}
The governing equations for the coupling between the solid mechanics field and the electrochemical field are
\begin{subequations}
\begin{alignat}{2}
    \mat{F}_\text{growth} &= \textbf{fn}(c) &&\quad \text{in} \ \Omega_\text{ed}, \\
    \mat{F}_\text{growth} &= \mat{1} &&\quad \text{in} \ \Omega \ \backslash \ \Omega_\text{ed}.
\end{alignat}
\end{subequations}
\subsection{A novel model to capture transport in coating layers}
Within the domain of the coating layer~$\Omega_\text{coat}$, we formulate assumptions that we consider as reasonable as they are based on a priori knowledge. With these assumptions, the governing equations in the coating layer are adapted.
\subsubsection{Assumptions in the coating layer}
\label{sec:coating_layer_assumptions}
We formulate two assumptions inside the domain of the coating layer
\begin{enumerate}
    \item One assumption for the electric field~$\vec{E} = - \nabla \Phi$ inside of the coating layer is formulated due to the geometric characteristic of the coating layer of having a small curvature and being thin compared to the other geometric domains. For an isotropic ionic conductivity of the coating material, we state that the electric field in the normal direction~$E_\text{n} = \vec{E} \cdot \vec{n}$ is constant along the normal direction of the coating layer.
    \item Another assumption is that the mechanical impact of the coating layer is neglected, and the coating layer is inseparably connected to the adjacent bulk domains. This implies that no strain of the coating layer in the normal direction occurs. Thus, we consider only strains in the tangential direction caused by the adjacent bulk domains. However, similar assumptions as for the electric field could be formulated for the strains and introduced to the equations of solid mechanics if the investigation focuses on the mechanical influence of the coating layer, e.g., to analyze damage of the material of the coating layer.
\end{enumerate}
With both assumptions, the governing equations in the coating layer for the conservation of mass and charge read
\begin{alignat}{2}
    -\kappa \Delta_\Gamma \Phi &= \frac{1}{t_\text{coat}}\frac{\eta}{r_\text{n}} + s_{\rho,  \text{coup}} &&\quad \text{on} \ \Gamma_\text{coat}, \\
    \left.\parder{c}{t}\right|_{\vec{X}} + c \ \nabla_\Gamma \cdot \dot{\vec{u}} &= s_{c,  \text{coup}} &&\quad \text{on} \ \Gamma_\text{coat},
\end{alignat}
where $\Delta_\Gamma \Phi$ denotes the Laplace operator, and $\nabla_\Gamma \cdot$ the divergence on a two-dimensional manifold $\Gamma_\text{coat}$ (see Appendix \ref{sec:curved_nabla} for a derivation). The term~$\frac{1}{t_\text{coat}}\frac{\eta}{r_\text{n}}$ originates from the constant of integration as we will show. The source term $s_{\rho, \text{coup}} = \frac{\vec{j}_\text{coup} \cdot \vec{n}}{t_\text{coat}}$ considers the coupling fluxes to the adjacent bulk domains. As common for ion conductors, the current density and the mass flux density are linked via~$\frac{t_+}{z F}$, such that $s_{c, \text{coup}} = \frac{t_+}{z F} \frac{\vec{j}_\text{coup} \cdot \vec{n}}{t_\text{coat}}$. The coupling flux density $\vec{j}_\text{coup}$ is described by a flux model as introduced later. The difference in potential across the coating layer is~$\eta = \Phi_\text{coat-el} - \Phi_\text{coat-ed}$ and the resistance in normal direction is~$r_\text{n} = \frac{t_\text{coat}}{\kappa}$.
\subsubsection{Range of validity of the assumption concerning the electric field}
We quantify the range of validity of the assumption by evaluating the electric field inside a coating layer of a cylindrical electrode with a fully resolved model. For this geometric setup, an analytic solution can be found if the equations are evaluated in cylindrical coordinates. We begin with analyzing the equation for the conservation of charge (\cref{eq:pot_solid_electrolyte}) with an isotropic ionic conductivity
\begin{equation}
    \label{eq:laplace_cylindrical_coordinates}
    \Delta \Phi = \frac{\partial^2\Phi}{\partial r^2} + \frac{1}{r} \parder{\Phi}{r} + \frac{1}{r^2} \frac{\partial^2 \Phi}{\partial \theta^2} + \frac{\partial^2 \Phi}{\partial z^2} = 0.
\end{equation}
If the first assumption is evaluated in cylindrical coordinates and the radial coordinate of the cylindrical coordinate system equals the normal direction of the coating layer, the first assumption reads $\frac{\partial^2\Phi}{\partial r^2}=0$. Integrating this constraint w.r.t. the radial coordinate leads to $\parder{\Phi}{r} = c_1(\theta, z)$. Substituting $\frac{\partial^2\Phi}{\partial r^2}=0$ and $\parder{\Phi}{r} = c_1(\theta, z)$ in \cref{eq:laplace_cylindrical_coordinates} gives
\begin{equation}
    \Delta \Phi = \frac{1}{r} c_1(\theta, z) + \frac{1}{r^2} \frac{\partial^2 \Phi}{\partial \theta^2} + \frac{\partial^2 \Phi}{\partial z^2} = \frac{1}{r} c_1(\theta, z) + \Delta_\Gamma \Phi = 0.
\end{equation}
From this, we can conclude that the assumption of the normal component of the electric field being constant in the normal direction is valid, if~$\frac{1}{r} = \text{const.} = \frac{1}{r_\text{in}} = \frac{1}{r_\text{in} + t_\text{coat}}$, with the inner radius of the coating layer~$r_\text{in}$ and the thickness of the coating layer~$t_\text{coat}$. This is profound, if~$r_\text{in}$ is either large (i.e. small curvatures of the coating layer) or the thickness of the coating layer~$t_\text{coat}$ is small. Generalized, this means that our assumption is profound if $r_\text{in} >> t_\text{coat}$. If this requirement is fulfilled, the conservation of charge is given by the Poisson equation $\Delta_\Gamma \Phi = c_1^*(\theta, z)$. As a consequence of the constitutive law for the current density, it follows that for the normal component of the current density in the coating layer: $\frac{i_\text{n}}{t_\text{coat}} = \frac{\kappa}{t_\text{coat}} E_n = \text{const} = \frac{\eta}{t_\text{coat}} \frac{\kappa}{t_\text{coat}} = \frac{1}{t_\text{coat}} \frac{\eta}{r_\text{n}}$. This allows to define $c_1^*(\theta, z) = \frac{1}{t_\text{coat}} \frac{\eta}{r_\text{n}}$, such that the equation for the conservation of charge in the coating layer is
\begin{equation}
    - \kappa \Delta_\Gamma \Phi = \frac{1}{t_\text{coat}}\frac{\eta}{r_\text{n}} + s_{\rho,  \text{coup}} \quad \text{on} \ \Gamma_\text{coat},
\end{equation}
as stated before.
\subsubsection{Geometric implication of the assumptions}
By not resolving the thickness of the coating layer geometrically but reducing it to a two-dimensional manifold, a gap in the three-dimensionally resolved geometry would occur. This gap can be assigned to the electrode domain ($\Gamma_\text{coat}$ is located at $\Gamma_\text{el-coat}$), to the electrolyte domain ($\Gamma_\text{coat}$ is located at $\Gamma_\text{coat-c}$), or left as a geometrical gap. Assigning the gap to the electrode domain modifies the capacity of the cell. Assigning it to the electrolyte influences the total resistance of the cell in general more, as usually, the ionic conductivity of the solid electrolyte is smaller compared to the electronic conductivity of the electrodes. Hence, keeping the gap would be the best option from an electrochemical perspective but would require specific features in the underlying code, like defining a consistent surface in the center of the gap for non-convex and non-steady differentiable surface curvatures, that are not difficult from a theoretical point of view but challenging w.r.t. implementations and are often not available. Thus, the gap is assigned to the electrolyte in this work to maintain the capacity of the battery cell. However, the novel approach can be used with all three possibilities, depending on the available code.
\subsection{Models for interface phenomena}
\label{sec:interface_models}
We distinguish between two types of interface models for coupling the different domains of ASSBs. One model type assumes a continuous flux density between two domains across the interface based on the difference in a driving potential. The other model type assumes continuity of a solution quantity on both sides of the interface.
\subsubsection{Models for the coupling flux densities}
\label{sec:coupling_flux}
Coupling equations of this type model the kinetics at an interface and relate the electric current density~$i$ and the mass flux density~$j$ across the interface with a difference in potential at both sides of the interface~$i = \text{fn}(\eta)$ and~$j = \text{fn}(\eta)$, with the overpotential~$\eta = \Phi_\text{ed} - \Phi_\text{el} - \Phi_0(c)$, where~$\Phi_0(c)$ denotes the equilibrium potential as a function of the concentration, and $\Phi_\text{ed}$ and $\Phi_\text{el}$ the electric potential at electrode and electrolyte sides of the interface, respectively. These types of constraints are applied as Robin-type interface conditions. \\
The conservation of both mass and charge requires the net sum of the flux densities at an interface between the domains~$\Omega_i$ and~$\Omega_j$ to be zero, such that $\vec{i}_i \cdot \vec{n}_i = i = - \vec{i}_j \cdot \vec{n}_j$ and $\vec{j}_i \cdot \vec{n}_i = j = - \vec{j}_j \cdot \vec{n}_j$.
\paragraph{Butler--Volmer interface kinetics}
The chemical reaction at the interface between electrolyte and electrodes is modeled by the Butler-Volmer kinetics, which links the overpotential~$\eta$ to the current density across the interface
\begin{equation}
i = i_0 \left[\text{exp} \left(\frac{\alpha_\text{a} F \eta}{R T}\right) - \text{exp} \left(\frac{-(1-\alpha_\text{a}) F \eta}{R T}\right) \right] \quad \text{on} \ \Gamma_\text{coat-c} \cup \Gamma_\text{a-el},
\end{equation}
with the temperature~$T$, which we consider constant throughout this work ($T=298 \ \text{K}$). As charge inside the solid electrolyte is carried by ions, we define $\vec{i}_\text{el} = \vec{j}_\text{el} z\frac{F}{t_+} \ \text{on} \ \Gamma_\text{coat-c} \cup \Gamma_\text{a-el}$. The reaction at the electrode side requires uncharged species and electrons in a ratio of~$z F$, such that the electric current density and the mass flux density are linked~$\vec{i}_\text{ed} = z F \vec{j}_\text{ed} \ \text{on} \ \Gamma_\text{coat-c} \cup \Gamma_\text{a-el}$.
\paragraph{Ohmic resistance and constant permeability}
The kinetics at other interfaces where no chemical charge transfer reactions occur are modeled by linear laws. At the interface between the current collectors and the electrodes, an Ohmic law with resistance~$r_\text{i}$ is given as
\begin{equation}
    i = \frac{\Phi_\text{cc} - \Phi_\text{ed}}{r_\text{i}} \quad \text{on} \ \Gamma_\text{cc-c} \cup \Gamma_\text{cc-a},
\end{equation}
and a mass flux density of~$j=0$ as the current collectors are impermeable for lithium (-ions). At the interface between the coating layer and the bulk electrolyte, a resistance for the flux density of ions is given as
\begin{equation}
    j = \frac{t_+}{z F} \frac{\Phi_\text{coat} - \Phi_\text{el}}{r_\text{i}} \quad \text{on} \ \Gamma_\text{el-coat}.
\end{equation}
Again, the electric current density and the mass flux density are linked $i=j z \frac{F}{t}$.
\paragraph{No interface flux}
There is no flux of charge $i=0$ and mass $j=0$ at the interfaces $\Gamma_\text{cc-el}$ and $\Gamma_\text{cc-coat}$ as the current collectors are impermeable for lithium-ions and the solid electrolyte, and the coating layer is electronically isolating.
\subsubsection{Models enforcing continuity}
\label{sec:pointwise_coupling}
The continuity of a quantity~$\Psi$ across an interface is enforced by constraints of the form
\begin{equation}
    \Psi_i = \Psi_j \quad \text{on} \ \Gamma_{i-j}.
\end{equation}
In this study, we enforce that the domains cannot separate, i.e. the displacements on both sides of the interface are equal
\begin{equation}
    \vec{u}_i = \vec{u}_j \quad \text{on} \ \Gamma_{i-j}.
\end{equation}
This means that for simplicity, we neglect the potential loss of contact between two domains due to tensile stresses or shear movements but will consider this in a future publication. We model continuity of the electric potential and of the concentration where two coating domains are adjacent
\begin{alignat}{2}
    \Phi_i &= \Phi_j &&\quad \text{on} \ \Gamma_{\text{coat,} i} \cap \Gamma_{\text{coat,} j}, \\
    c_i &= c_j &&\quad \text{on} \ \Gamma_{\text{coat,} i} \cap \Gamma_{\text{coat,} j},
\end{alignat}
with $\Gamma_{\text{coat,} i}$ denoting the coating layer $i$. To add these types of constraints to the system of equations, Lagrange multipliers $\lambda_\Psi$ are introduced
\begin{equation}
    \label{eq:energy_constraints}
    W_\text{constr} = \int_\Gamma \left[ \vec{\lambda}_{\vec{u}}^\text{T} \left( \vec{u}_\text{m} - \vec{u}_\text{s} \right) \right] \text{d} \Gamma + \int_{\partial\Gamma} \left[ \lambda_c \left( c_\text{m} - c_\text{s} \right) \right] \text{d} \partial \Gamma + \int_{\partial\Gamma} \left[ \lambda_\Phi \left( \Phi_\text{m} - \Phi_\text{s} \right) \right] \text{d} \partial \Gamma.
\end{equation}
The indices 'm' and 's' denote the adjacent, coupled geometric entities with 'master' and 'slave' as commonly done, and $\partial \Gamma$ represents the intersection line of two two-dimensional manifolds.
\subsection{Boundary and initial conditions}
Proper boundary and initial conditions are required to obtain a well-posed system of equations. Both the flux of mass and charge are prevented across artificial model boundaries by
\begin{alignat}{2}
    \vec{j} \cdot \vec{n} &= 0 \quad && \text{on} \ \Gamma_\text{cut}, \\
    \vec{i} \cdot \vec{n} &= 0 \quad && \text{on} \ \Gamma_\text{cut}.
\end{alignat}
Furthermore, a flux of mass across the outer boundaries of the current collectors is not possible
\begin{equation}
    \vec{j} \cdot \vec{n} = 0 \quad \text{on} \ \Gamma_\text{cc,a-o} \cup \Gamma_\text{cc,c-o}.
\end{equation}
The mass and charge transfer inside the battery cell is caused by a fixed value of the electric potential at the anode side current collector
\begin{equation}
    \Phi = 0 \quad \text{on} \ \Gamma_\text{cc,a-o},
\end{equation}
in combination with an electric current density at the cathode side current collector
\begin{equation}
    -\vec{i} \cdot \vec{n} = \hat{i} \quad \text{on} \ \Gamma_\text{cc,c-o}.
\end{equation}
Initially, values for the concentrations in the electrodes and the solid electrolyte are set to
\begin{equation}
    c(t=0) =
    \begin{cases}
        c_{0,\text{a}} \quad & \text{in} \ \Omega_\text{a} \\
        c_{0,\text{c}} \quad & \text{in} \ \Omega_\text{c} \\
        c_{0,\text{el}} \quad & \text{in} \ \Omega_\text{el} \\
        c_{0,\text{coat}} \quad & \text{in} \ \Omega_\text{coat}
    \end{cases}
    .
\end{equation}
In this work, we assume a perfectly stiff mechanical housing of the battery cell, meaning that all displacements normal to outer boundaries vanish
\begin{equation}
    \vec{u} \cdot \vec{n} = 0 \quad \text{on} \ \Gamma_\text{cc-o}.
\end{equation}
Equivalently, the displacement in the normal direction is set to zero on the boundaries~$\Gamma_\text{cut}$. Finally, zero displacements and velocities are assumed in all domains at the initial state
\begin{equation}
    \vec{u}(t=0) = \vec{\dot{u}}(t=0) = \vec{0} \quad \text{in} \ \Omega_0.
\end{equation}

%% file: figures/geometry.pdf_tex
%% Creator: Inkscape inkscape 0.92.5, www.inkscape.org
%% Accompanies image file 'geometry.pdf' (pdf, eps, ps)
\begingroup%
  \makeatletter%
  \providecommand\color[2][]{%
    \errmessage{(Inkscape) Color is used for the text in Inkscape, but the package 'color.sty' is not loaded}%
    \renewcommand\color[2][]{}%
  }%
  \providecommand\transparent[1]{%
    \errmessage{(Inkscape) Transparency is used (non-zero) for the text in Inkscape, but the package 'transparent.sty' is not loaded}%
    \renewcommand\transparent[1]{}%
  }%
  \providecommand\rotatebox[2]{#2}%
  \newcommand*\fsize{\dimexpr\f@size pt\relax}%
  \newcommand*\lineheight[1]{\fontsize{\fsize}{#1\fsize}\selectfont}%
  \ifx\svgwidth\undefined%
    \setlength{\unitlength}{194.96358959bp}%
    \ifx\svgscale\undefined%
      \relax%
    \else%
      \setlength{\unitlength}{\unitlength * \real{\svgscale}}%
    \fi%
  \else%
    \setlength{\unitlength}{\svgwidth}%
  \fi%
  \global\let\svgwidth\undefined%
  \global\let\svgscale\undefined%
  \makeatother%
  \begin{picture}(1,0.32288673)%
    \lineheight{1}%
    \setlength\tabcolsep{0pt}%
    \put(0,0){\includegraphics[width=\unitlength,page=1]{geometry.pdf}}%
    \put(0.26077883,0.14984842){\color[rgb]{0,0,0}\makebox(0,0)[t]{\lineheight{1.25}\smash{\begin{tabular}[t]{c}$\Omega_\text{a}$\end{tabular}}}}%
    \put(0.13847681,0.15009035){\color[rgb]{0,0,0}\makebox(0,0)[t]{\lineheight{1.25}\smash{\begin{tabular}[t]{c}$\Omega_\text{cc,a}$\end{tabular}}}}%
    \put(0.55481439,0.14950499){\color[rgb]{0,0,0}\makebox(0,0)[t]{\lineheight{1.25}\smash{\begin{tabular}[t]{c}$\Omega_\text{el}$\end{tabular}}}}%
    \put(0.75130084,0.21581958){\color[rgb]{0,0,0}\makebox(0,0)[t]{\lineheight{1.25}\smash{\begin{tabular}[t]{c}$\Omega_\text{c}$\end{tabular}}}}%
    \put(0.86578887,0.16743416){\color[rgb]{0,0,0}\makebox(0,0)[t]{\lineheight{1.25}\smash{\begin{tabular}[t]{c}$\Omega_\text{cc,c}$\end{tabular}}}}%
    \put(0.43159578,0.30978137){\color[rgb]{0,0,0}\makebox(0,0)[t]{\lineheight{1.25}\smash{\begin{tabular}[t]{c}$\Gamma_\text{cut}$\end{tabular}}}}%
    \put(0.82385754,0.30978137){\color[rgb]{0,0,0}\makebox(0,0)[t]{\lineheight{1.25}\smash{\begin{tabular}[t]{c}$\color{cccc}\Gamma_\text{c-cc}$\end{tabular}}}}%
    \put(0.82361759,-0.00281482){\color[rgb]{0,0,0}\makebox(0,0)[t]{\lineheight{1.25}\smash{\begin{tabular}[t]{c}$\color{elccc}\Gamma_\text{cc-el}$\end{tabular}}}}%
    \put(0.34137748,-0.00281485){\color[rgb]{0,0,0}\makebox(0,0)[t]{\lineheight{1.25}\smash{\begin{tabular}[t]{c}$\color{ael}\Gamma_\text{a-el}$\end{tabular}}}}%
    \put(0.90374304,-0.00281485){\color[rgb]{0,0,0}\makebox(0,0)[t]{\lineheight{1.25}\smash{\begin{tabular}[t]{c}$\color{ccco}\Gamma_\text{cc,c-o}$\end{tabular}}}}%
    \put(0.17927902,-0.00281485){\color[rgb]{0,0,0}\makebox(0,0)[t]{\lineheight{1.25}\smash{\begin{tabular}[t]{c}$\color{ccaa}\Gamma_\text{cc-a}$\end{tabular}}}}%
    \put(0.09690311,-0.00281485){\color[rgb]{0,0,0}\makebox(0,0)[t]{\lineheight{1.25}\smash{\begin{tabular}[t]{c}$\color{ccao}\Gamma_\text{cc,a-o}$\end{tabular}}}}%
    \put(0.78464735,0.09974902){\color[rgb]{0,0,0}\makebox(0,0)[t]{\lineheight{1.25}\smash{\begin{tabular}[t]{c}$\color{elcoat}\Gamma_\text{el-coat}$\end{tabular}}}}%
    \put(0,0){\includegraphics[width=\unitlength,page=2]{geometry.pdf}}%
    \put(0.69470399,0.17764712){\color[rgb]{0,0,0}\rotatebox{-45}{\makebox(0,0)[t]{\lineheight{1.25}\smash{\begin{tabular}[t]{c}$\Omega_\text{coat}$\end{tabular}}}}}%
    \put(6.61112308,-1.0969693){\color[rgb]{0,0,0}\makebox(0,0)[lt]{\begin{minipage}{1.6097149\unitlength}\raggedright \end{minipage}}}%
    \put(0.66846071,0.30985864){\color[rgb]{0,0,0}\makebox(0,0)[t]{\lineheight{1.25}\smash{\begin{tabular}[t]{c}$\color{coatc}\Gamma_\text{coat-c}$\end{tabular}}}}%
    \put(0,0){\includegraphics[width=\unitlength,page=3]{geometry.pdf}}%
    \put(0.86590601,0.12124136){\color[rgb]{0,0,0}\makebox(0,0)[t]{\lineheight{1.25}\smash{\begin{tabular}[t]{c}$\Gamma_\text{cc-coat}$\end{tabular}}}}%
    \put(0,0){\includegraphics[width=\unitlength,page=4]{geometry.pdf}}%
    \put(0.64591301,0.27905379){\color[rgb]{0,0,0}\makebox(0,0)[lt]{\lineheight{1.25}\smash{\begin{tabular}[t]{l}$t_1$\end{tabular}}}}%
    \put(0.63016211,0.213399){\color[rgb]{0,0,0}\makebox(0,0)[lt]{\lineheight{1.25}\smash{\begin{tabular}[t]{l}$n$\end{tabular}}}}%
  \end{picture}%
\endgroup%

%% file: chapters/chapter_3.tex
The governing equations are discretized in time, using the one-step-theta method, and in space, using the finite-element method. The algebraic, nonlinear system is iteratively solved using the Newton-Raphson scheme, and the resulting linear system of equations is solved using tailored iterative linear solvers. We use~$\Psi$ as an exemplary variable in the following.
\subsection{Discretization in time}
The one-step-theta method is used throughout this work to discretize the equations in time. It is applied to discretize partial differential equations of first order of the form $\parder{\Psi}{t} = \text{fn}(\Psi, \vec{x}, t)$ in time for $t \in [t_0, t_\text{end}]$ with the scheme
\begin{equation}
    \parder{\Psi}{t} \approx \frac{\Psi^{n+1} - \Psi^n}{\Delta t} = \theta \text{fn}(\Psi^{n+1}, \vec{x}^{n+1}, t^{n+1}) + (1- \theta)\text{fn}(\Psi^n, \vec{x}^n, t^n),
\end{equation}
with the size of the time step~$\Delta t$. $\Psi^{n+1}$ and $\Psi^n$ denote the discrete states of~$\Psi$ at different time steps~$t^{n+1}$ and~$t^n$, respectively. Equivalently, $\vec{x}$ is evaluated at~$t^{n+1}$ and~$t^n$. In the following, we omit the superscript~$n+1$ for readability.
\subsection{Weak formulation of the ASSB model}
As an essential step to allow for spatial discretization using the finite-element method, the time discrete equations are reformulated to the weak formulation by multiplying the balance equations, Neumann boundary conditions, and Robin interface conditions by an arbitrary test function~$\delta g_\Psi$, integrating over the domains or boundaries, respectively, and applying Gau{\ss} divergence theorem. The contribution of the constraints to the weak formulation that enforce continuity is computed by the variation of \cref{eq:energy_constraints}
\begin{multline}
    \delta W_\text{constr} = \int_\Gamma \left[ \delta \vec{\lambda}_{\vec{u}}^\text{T} \left( \vec{u}_\text{m} - \vec{u}_\text{s} \right) + \left( \delta \vec{u}_\text{m} - \delta \vec{u}_\text{s} \right)^\text{T} \vec{\lambda}_{\vec{u}} \right] \text{d} \Gamma + \\ \int_{\partial \Gamma} \left[ \delta \lambda_c \left( c_\text{m} - c_\text{s} \right) + \left( \delta c_\text{m} - \delta c_\text{s} \right) \lambda_c \right] \text{d} \partial \Gamma + \int_{\partial \Gamma} \left[ \delta \lambda_\Phi \left( \Phi_\text{m} - \Phi_\text{s} \right) + \left( \delta  \Phi_\text{m} - \delta \Phi_\text{s} \right) \lambda_\Phi \right] \text{d} \partial \Gamma,
\end{multline}
with the boundary of a two-dimensional manifold~$\partial \Gamma$. Finally, all contributions to the weak formulation (from electrochemistry~$\delta W_\text{elch}$, solid mechanics~$\delta W_\text{mech}$, and the constraints~$\delta W_\text{constr}$) are summed up, as done in our previous work~\cite{Schmidt2022}
\begin{equation}
    \delta W = \delta W_\text{elch} + \delta W_\text{mech} + \delta W_\text{constr}.
\end{equation}
Equivalent to solving the governing equations is to find the solution of $\delta W = 0$.
\subsection{Discretization in space}
The finite-element method is used to discretize the equations of the weak formulation in space. The domains are split into finite elements, and thus, the integration over the domain is divided into the sum of all elements and the integration over these elements. Primary variables, test functions, and the shape of the elements are approximated by Lagrangian polynomials multiplied by the nodal values~$\Psi=\mat{N}\vec{\Psi}$ within the elements. Throughout this work, linear polynomials are used for the shape functions~$\mat{N}$. The resulting expression can be reorganized as
\begin{multline}
    \delta W = {\delta \vec{g}_c^\text{bulk}}^\text{T} \left(\Rcbulknp + \Rcbulksurfnp \right) + {\delta \vec{g}_\Phi^\text{bulk}}^\text{T} \left(\Rpotbulknp + \Rpotbulksurfnp \right) + \\ {\delta \vec{g}_c^\text{manif}}^\text{T} \left(\Rcsurfnp - \frac{1}{t_\text{coat}} \Rcbulksurfnp \right) + {\delta \vec{g}_\Phi^\text{manif}}^\text{T} \left(\Rpotsurfnp - \frac{1}{t_\text{coat}} \Rpotbulksurfnp \right) + \delta \vec{g}_{\vec{u}}^\text{T} \Rudefnp + \delta \vec{g}_{\lambda_i}^\text{T} \vec{R}_{\lambda_i} = 0,
\end{multline}
where $\mat{R}_{\vec{\Psi}}^{\text{bulk-manif}}$ denotes the residual of the coupling constraints between the bulk and the two-dimensional manifolds and $\mat{R}_{\vec{\Psi}}^{*}$ denotes the residuals of the different domains with the unknowns $\vec{\Psi} = \{\vec{u}, \vec{\lambda}_\Psi, \vec{c}, \vec{\Phi} \}$. The values for the test functions~$\delta \vec{g}_\Psi$ are arbitrary, such that all residuals have to be individually zero
\begin{equation}
    \label{eq:nonlinear_system}
    \begingroup
    \renewcommand*{\arraystretch}{1.6}
    \begin{bmatrix}
        \Rudefnp (\udefnp, \lambdadefnp, \cbulknp) \\
        \Rlambdadefnp(\udefnp) \\
        \Rcbulknp (\udefnp, \cbulknp, \potbulknp) + \Rcbulksurfnp (\udefnp, \cbulknp, \potbulknp, \csurfnp, \potsurfnp)\\
        \Rpotbulknp (\udefnp, \cbulknp, \potbulknp) + \Rpotbulksurfnp (\udefnp, \cbulknp, \potbulknp, \csurfnp, \potsurfnp)\\
        \Rcsurfnp (\udefnp, \csurfnp, \potsurfnp, \lambdacsurfnp) - \frac{1}{t_\text{coat}} \Rcbulksurfnp (\udefnp, \cbulknp, \potbulknp, \csurfnp, \potsurfnp, \lambdacsurfnp) \\
        \Rpotsurfnp (\udefnp, \csurfnp, \potsurfnp, \lambdapotsurfnp) - \frac{1}{t_\text{coat}} \Rpotbulksurfnp (\udefnp, \cbulknp, \potbulknp, \csurfnp, \potsurfnp, \lambdapotsurfnp) \\
        \Rlambdacsurfnp (\csurfnp) \\
        \Rlambdapotsurfnp (\potsurfnp)
    \end{bmatrix}
    =
    \vec{0}.
    \endgroup
\end{equation}
\subsection{Solution of the nonlinear algebraic system of equations}
The nonlinear system of equations (\cref{eq:nonlinear_system}) is iteratively solved using the Newton-Raphson scheme $\vec{\Psi}_{i+1} = \vec{\Psi}_i - \Delta \vec{\Psi}_{i+1}$, with
\begin{equation}
    \label{eq:linearized_system}
    \Delta \vec{\Psi}_{i+1} = \left(\left.\parder{\vec{R}}{\vec{\Psi}}\right|_{i}\right)^{-1} \vec{R}(\vec{\Psi}_i),
\end{equation}
for the unknowns~$\vec{\Psi} = \{\vec{u}, \vec{\lambda}_\Psi, \vec{c}, \vec{\Phi} \}$.
\subsection{Static condensation of the linearized system of equations}
Static condensation is used to remove the Lagrange multipliers and the slave side values from the linearized system of equations (\cref{eq:linearized_system}). This also removes the saddle-point structure of the system of equations (i.e. zero entries on the main diagonal).
\begin{comment}
%
\begin{figure}[ht]
    \centering
    \def\svgwidth{0.5\textwidth}
    \import{figures/}{manifold_mesh_tying.pdf_tex}
    \caption{Schematic sketch of mesh tying between two-dimensional manifolds (purple) and domains (black).}
    \label{fig:mesh_tying}
\end{figure}
\end{comment}
%
We divide the vectors $\udefnp$, $\csurfnp$, and $\potsurfnp$ into vectors with unknowns on the interior, master, and slave side. For brevity, we add a tilde to the indices in the submatrices of the two-dimensional manifolds~$\tilde{*}$ and a hat to indices in the bulk~$\hat{*}$ of the condensed system of equations
\begingroup
\small
\begin{equation}
    \label{eq:condensed_linearized_system}
    \renewcommand*{\arraystretch}{1.6}
    \left.
    \begin{bmatrix}
        \mat{K}_{\hat{u},\hat{u}} & \mat{K}_{\hat{u}, \hat{c}} & \mat{0} & \mat{0} & \mat{0} \\
        \mat{K}_{\hat{c}, \hat{u}} & \mat{K}_{\hat{c}, \hat{c}} & \mat{K}_{\hat{c}, \hat{\phi}} & \mat{K}_{\hat{c}, \tilde{c}} & \mat{K}_{\hat{c}, \tilde{\phi}} \\
        \mat{K}_{\hat{\phi}, \hat{u}} & \mat{K}_{\hat{\phi}, \hat{c}} & \mat{K}_{\hat{\phi}, \hat{\phi}} & \mat{K}_{\hat{\phi}, \tilde{c}} & \mat{K}_{\hat{\phi}, \tilde{\phi}} \\
        \mat{K}_{\tilde{c}, \hat{u}} & \mat{K}_{\tilde{c}, \hat{c}} & \mat{K}_{\tilde{c}, \hat{\phi}} & \mat{K}_{\tilde{c}, \tilde{c}} & \mat{K}_{\tilde{c}, \tilde{\phi}} \\
        \mat{K}_{\tilde{\phi}, \hat{u}} & \mat{K}_{\tilde{\phi}, \hat{c}} & \mat{K}_{\tilde{\phi}, \hat{\phi}} & \mat{K}_{\tilde{\phi}, \tilde{c}} & \mat{K}_{\tilde{\phi}, \tilde{\phi}}
    \end{bmatrix}
    \right|_j
    \left.
    \begin{bmatrix}
        \Delta \udefnp \\
        \Delta \cbulknp \\
        \Delta \potbulknp \\
        \Delta \csurfnp \\
        \Delta \potsurfnp
    \end{bmatrix}
    \right|_{j+1}
    =
    -
    \left.
    \begin{bmatrix}
        \Rudefnp \\
        \Rcbulknp + \Rcbulksurfnp \\
        \Rpotbulknp + \Rpotbulksurfnp \\
        \Rcsurfnp - \frac{1}{t_\text{coat}} \Rcbulksurfnp  \\
        \Rpotsurfnp  - \frac{1}{t_\text{coat}} \Rpotbulksurfnp \\
    \end{bmatrix} 
    \right|_j .
\end{equation}
\endgroup
\subsection{Solution of the linearized system of equations}
We decided to monolithically \cite{Verdugo2016} solve the linearized system of equations (\cref{eq:condensed_linearized_system}) due to the superior robustness and efficiency of the solution scheme compared to other strategies like partitioned coupling. As this linear system is, in general, poorly conditioned due to the different physical fields, different geometric entities, and additional off-diagonal contributions, tailored solvers using a combination of Block-Gau\ss-Seidel and Algebraic-Multigrid preconditioners~\cite{Fang2019} are used to make iterative solvers applicable and thus the solution of large and realistic systems feasible.

%% file: chapters/chapter_4.tex
The presented model and the numerical approach are implemented in the software project \textit{BACI}~\cite{BACI}. In the following, we show the conservation of mass for the novel model, verify the assumptions, and apply the model to geometrically complex scenarios to show its capabilities.
\subsection{Verification of the outlined model}
First, we show the conservation of mass of the model for a deforming geometry and quantify the numerical error in terms of temporal convergence. Afterwards, the assumptions we made in the coating layer are justified, and the results computed with the presented approach are compared to the results computed with a three-dimensionally resolved coating layer.
\subsubsection{Conservation of mass and error convergence}
\begin{figure}[ht]
    \begin{floatrow}
    \ffigbox{
        \def\svgwidth{0.4\textwidth}
        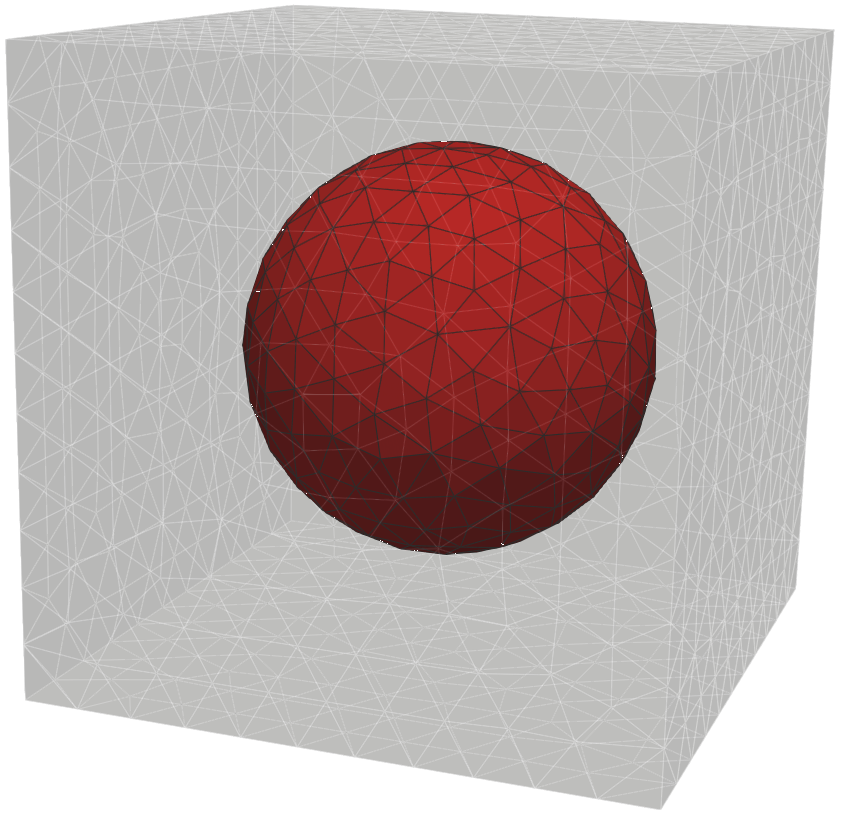
    }{
        \caption{Geometric representation of a sphere located in the center of a cube. Conservation of mass is enforced on the surface of the sphere.}
        \label{fig:geometryRepresentationSphere}
    }
    \capbtabbox{
        \input{tables/test_setup_sphere.tex}
    }{
        \caption{Geometric parameters for testing the conservation of mass.}
        \label{table:test_setup_sphere}
    }
    \end{floatrow}
\end{figure}

\begin{figure}[ht]
    \centering
    \input{figures/temporal_convergence_concentration_sphere.tikz}
    \caption{Convergence of the maximal value of L2-norm of error of the concentration for different time step sizes.}
    \label{fig:temporal_convergence_concentration_sphere}
\end{figure}
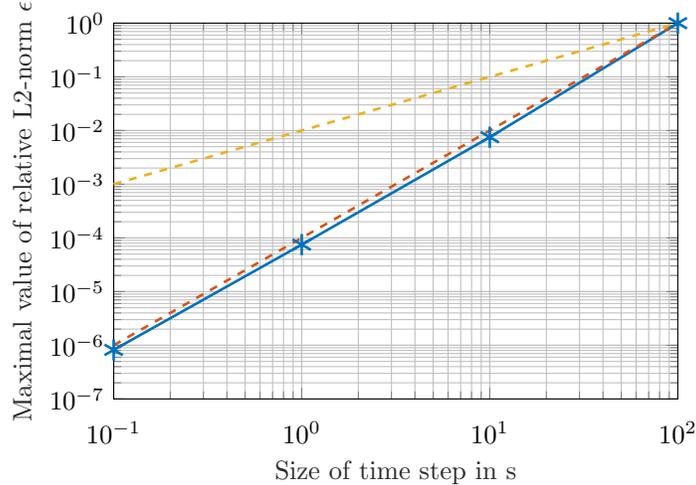

The model is tested for the conservation of mass by comparison of the results with an analytic solution. Therefore, a sphere is located in the center of a deforming cube (see \cref{fig:geometryRepresentationSphere} and \cref{table:test_setup_sphere}).
The surface of the sphere represents the two-dimensional manifold where the conservation equations for charge and mass are solved. An initial concentration~$c_\text{surf,0} = 1 \ \frac{\text{mol}}{\text{m}^3}$ is set on the surface without further external fluxes. The electric potential is fixed to $\Phi=0 \ \text{V}$ in all domains. An isotropic deformation $\mat{F} = \mat{1} (1 + \frac{t}{100 \ \text{s}})$ for t=[0 \ \text{s}, 100 \ \text{s}] is applied while rigid body motions are supressed. For this simple geometry and deformation state, an analytic expression for the concentration on the surface depending on the deformation is given using that the total mass remains constant $m = c_\text{ana}(t) A(t) t_\text{coat}  = c_0 A_0 t_\text{coat} = \text{const.}$, with the current surface area~$A(t)$ and the initial surface area $A_0 = A(t=0)$. The temporal development of the surface area is determined by the deformation gradient~$\mat{F}$, $A(t) = A_0 \left(\text{det}(\mat{F})\right)^\frac{2}{3}$. This leads to an analytic expression for the concentration 
\begin{equation}
    c_\text{ana}(t) = \frac{c_0}{(\text{det}(\mat{F}))^\frac{2}{3}}.
\end{equation}
Now, a temporal convergence study with $\Delta t = [0.1 \ \text{s}, 1 \ \text{s}, 10 \ \text{s}, 100 \ \text{s}]$ is performed by computing the relative L2-norm of the deviation of the concentration computed with the novel model from the analytically computed concentration
\begin{equation}
    \epsilon(t) = \frac{\sqrt{\int_{\Gamma_\text{coat}} \left[c_\text{ana}(t) - c(\vec{x},t) \right]^2 \ \text{d}\Gamma}}{\sqrt{\int_{\Gamma_\text{coat}} c_\text{ana}(t)^2 \ \text{d}\Gamma}}.
\end{equation}
In \cref{fig:temporal_convergence_concentration_sphere}, the maximal error $\epsilon_\text{max} = \text{max}(\epsilon(t))$ is shown for different sizes of the time step.
The dashed lines indicate the slope of linear and quadratic convergence, respectively. It is observable that the convergence rate of the error of the concentration converges quadratically w.r.t. the size of the time step, which can be expected from the one-step-theta method with~$\theta=0.5$.
\subsubsection{Justification of the assumptions by analyzing the solution of a geometrically resolved model}
\label{sec:results_verification}
\begin{figure}[ht]
    \begin{floatrow}
    \ffigbox{
        \def\svgwidth{0.4\textwidth}
        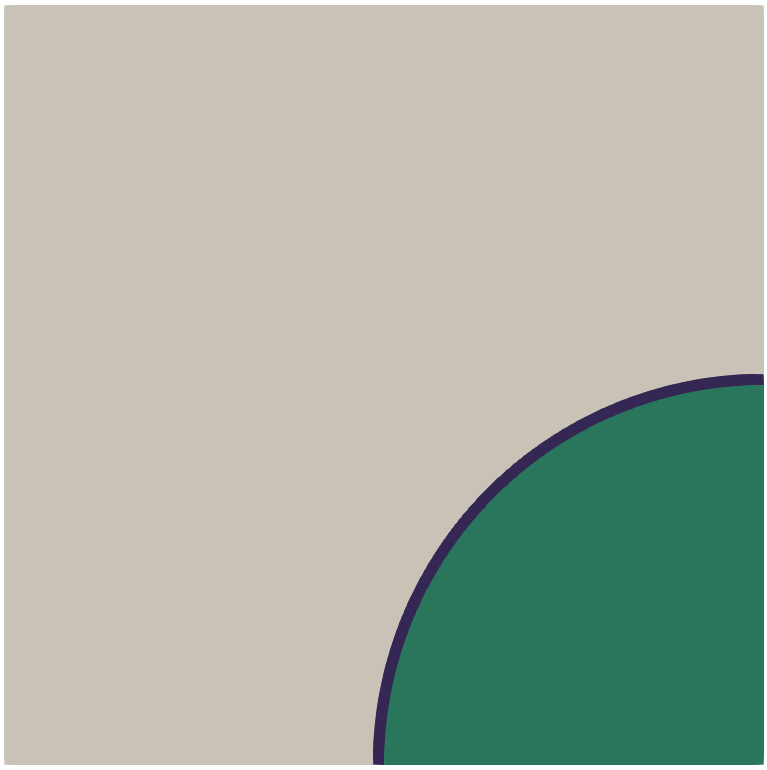
    }{
        \caption{Geometric representation of the geometry including the normal coordinate~$n$ and the tangential coordinate~$t_1$ along the coating layer.}
        \label{fig:comparison_geometry}
    }
    \capbtabbox{
        \input{tables/parameters_comparison_resolved.tex}
    }{
        \caption{Geometric parameters for comparing the resolved and the proposed model.}
        \label{table:parameters_comparison_resolved}
    }
    \end{floatrow}
\end{figure}

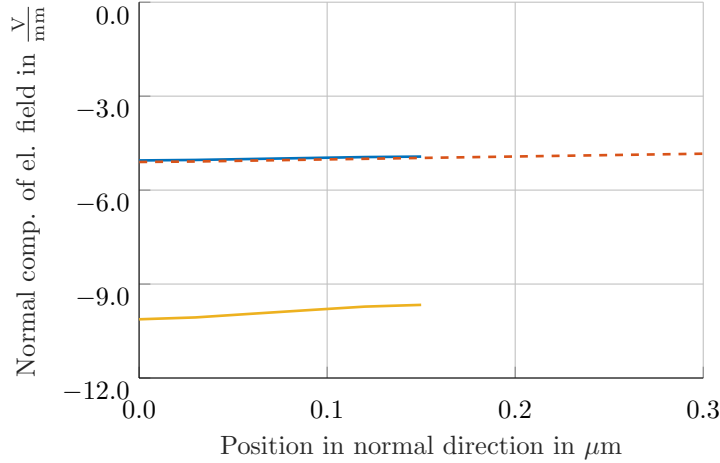
\begin{figure}[ht]
    \centering
    \input{figures/squared_geometry_standard_cond_radial_gradient_plot.tikz}
    \caption{Normal component of the electric field in the normal direction. The color of the lines indicates the geometric setup (blue: unmodified geometry, yellow: larger curvature, red: larger coating thickness).}
    \label{fig:squared_geometry_standard_cond_radial_gradient_plot}
\end{figure}

\begin{figure}[ht]
    \centering
    \input{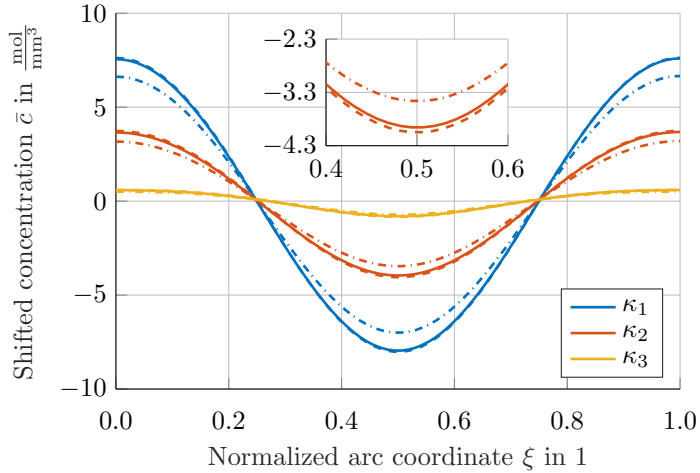}
    \caption{Comparison of the concentration in the electrode along the tangential coordinate~$t_1$ computed with the resolved model (solid lines) and with the outlined model (dashed lines: adding the thickness to the electrode, dash-dotted lines: adding the thickness to electrolyte) for different values of the ionic conductivity of the coating layer.}
    \label{fig:comparison_concentration_arc_length}
\end{figure}

We want to analyze the solution of a fully resolved model to justify the assumption we made and to compare the solution with the solution of the novel model. Therefore, a simple geometry is chosen that features non-planar curvatures of the coating layer and still can be three-dimensionally resolved: a cylindrical electrode is embedded into the solid electrolyte, and a thin coating layer is added on the surface of the electrode. By making use of the symmetry of the geometry, the size of the computational domain is reduced, i.e. reducing it to a quasi-two-dimensional geometry and considering only a quarter of the squared geometry (\cref{fig:comparison_geometry} and \cref{table:parameters_comparison_resolved}).
An electric current density $i=1 \frac{\text{A}}{\text{m}^2}$ is applied to the top and left sides of the electrolyte domain, and a constant electric potential $\Phi=0 \text{V}$ is set to the entire domain of the cylindrical electrode. The initial concentrations are $c_{0,\text{ed}} = 40000 \frac{\text{mol}}{\text{m}^3}$ in the electrode and $c_{0,\text{el}} = c_{0,\text{coat}} = 1200 \frac{\text{mol}}{\text{m}^3}$ in the solid electrolyte and the coating layer. The material parameters are summarized in \cref{table:material_parameters_validation}.
\paragraph{Justification of the assumption of a constant electric field in the normal direction along the normal direction}\mbox{}\\
We made the assumption that the normal component of the electric field is constant along the normal direction of the coating layer. For the justification of this assumption, the electric field is computed for the chosen setup with a geometrically fully resolved coating layer and analyzed at~$t=300 \ \text{s}$ when the gradient of the concentration in the electrode has reached the steady state. In \cref{fig:squared_geometry_standard_cond_radial_gradient_plot}, the normal component of the electric field is plotted along the normal direction at $\alpha = 45 \degree$ (see \cref{fig:comparison_geometry}) to justify the assumption that the normal component of the electric field is constant along the normal direction. In two other setups, the thickness of the coating layer is enlarged ($t_\text{coat} = 300 \ \text{nm}$, red line in \cref{fig:squared_geometry_standard_cond_radial_gradient_plot}) and the radius of the electrode is reduced ($r=2.5 \ \mu \text{m}$, yellow line in \cref{fig:squared_geometry_standard_cond_radial_gradient_plot}) to investigate the influence of the curvature and the thickness of the coating layer on the quality of the assumption.
It can be seen that the assumption of the normal component of the electric field being constant in the normal direction is profound, i.e. the lines are in a reasonable approximation constant. We define a measure to quantify the deviation from the assumption as the ratio of the difference of the minimal and the maximal value of the normal component of the electric field at $\alpha = 45\degree$ and the maximal value of the normal component of the electric field $\epsilon = \frac{\text{max}\left(E_\text{n}(\alpha = 45\degree)\right) - \text{min}\left(E_\text{n}(\alpha = 45\degree)\right)}{|\text{max}(E_\text{n}(\alpha = 45\degree))|}$. The deviation for the unmodified setup is~$\epsilon_\text{unmod} = 0.025$, which shows that the change of the normal component of the electric field in the normal direction is small compared to its absolute value and thus, our assumption is justified.\\
Furthermore, we see that reducing the radius of the cylindrical electrode (i.e. increasing the curvature of the coating layer) leads to an increased deviation from the assumption of~$\epsilon_\text{curv} = 0.048$, and increasing the thickness of the coating layer leads to an increased deviation of~$\epsilon_\text{thick}=0.055$. This means that our assumption is justified for small thicknesses and small curvatures of the coating layer. The deviation from the assumption increases for larger curvatures and larger thicknesses of the coating layer, as already stated before, but is still sufficiently small for realistic geometries.
\paragraph{Comparison of the results of the resolved and the presented model}\mbox{}\\
We compare the results of the presented model with the results of a model which three-dimensionally resolves the coating layer for the chosen setup. In one computational setup, the coating layer is geometrically resolved, and in another setup, the coating layer is captured with the presented model. Various simulations are performed to identify differences between the fully resolved model and the presented model, with the following parameters being varied:
\begin{itemize}
    \item Variation of the ionic conductivity of the coating layer: $\kappa_1=5 \cdot 10^{-4} \frac{\text{S}}{\text{m}}$, $\kappa_2=5 \cdot 10^{-3} \frac{\text{S}}{\text{m}}$, and $\kappa_3=5 \cdot 10^{-2} \frac{\text{S}}{\text{m}}$.
    \item Assignment of the gap due to the geometric reduction of the coating layer to the electrolyte.
    \item Assignment of the gap due to the geometric reduction of the coating layer to the electrode.
\end{itemize}
We formulate expectations for the concentration distribution on the surface of the electrode based on the knowledge that the concentration in the electrode will decrease due to the applied external discharge current, and the total exchanged charge must be equal for all simulations:
\begin{itemize}
    \item Increasing the ionic conductivity of the coating layer leads to a more homogeneous distribution of the concentration at the surface of the cathode due to favored conduction paths in the coating layer compared to conduction paths in the bulk electrolyte.
    \item Assigning the additional thickness to the electrolyte increases the resistance in the conduction paths in the electrolyte due to changes in the geometry of the electrolyte. The resistance scales with $l_\text{v}$ and $l_\text{c}$ (see \cref{fig:comparison_geometry}) at the vertices and at the center, respectively. Adding the thickness $t_\text{coat}$ to the electrolyte changes the ratio of the lengths $\frac{l_\text{v}}{l_\text{c}} > \frac{l_\text{v}+t_\text{coat}}{l_\text{c}+t_\text{coat}} > 1$, and thus a more homogenous resistance and a more homogenized concentration at the surface of the electrode are expected.
    \item Assigning the additional thickness to the electrode illustrates the influence of the assumption of the normal component of the electric field being constant along the normal direction on the concentration in the electrode. As already shown in \cref{fig:squared_geometry_standard_cond_radial_gradient_plot}, the deviation from the assumption of a constant electric field in the normal direction along the normal direction increases towards the vertices of the surface of the electrode. This has to be compensated by enlarged, unfavored conduction paths in the electrolyte domain (see \cref{fig:deviation_conduction_path}) which becomes more prominent towards the vertices. Thus, the flux density reduces there, and subsequently, the concentration at the vertices remains higher compared to the resolved model.
    \begin{figure}[ht]
        \def\svgwidth{0.3\textwidth}
        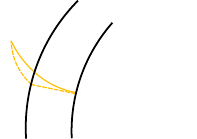
        \caption{Schematic deviation of the actual conduction path (solid line) in a coating layer and the attached solid electrolyte from point $A$ to point $B$ and the conduction path with the assumption of the normal component of the electric field being constant in the normal direction (dotted line) in the coating layer. The assumption leads to an enlargement of the total conduction path.}
        \label{fig:deviation_conduction_path}
    \end{figure}
\end{itemize}
In realistic setups, these three effects are always superimposed, such that a clear distinction is only possible in an academic setup like this. In \cref{fig:comparison_concentration_arc_length}, the concentration in the cylindrical electrode along the tangential coordinate~$t_1$ computed with the resolved and the novel model after a duration of~$t_\text{max} = 300 \ \text{s}$ is compared. Again, the point in time is chosen such that the gradient in the concentration has reached a steady state.
The solid lines represent the results from the resolved simulation, the dashed lines represent the results of the simulation where the additional thickness is assigned to the electrolyte, the dash-dotted lines represent the results where the thickness is assigned to the electrode, and the color codes refer to different ionic conductivities of the coating layer. To allow for comparison of the different geometries (i.e. the assignment of the gap to the electrode or the electrolyte), the arc length is normalized $\xi = \frac{t_1}{\text{max}(t_1)}$ and the averaged concentration on the surface is subtracted $\bar{c} = c - c_\text{avg}$.\\
The expected trends are observed in \cref{fig:comparison_concentration_arc_length}: An increased ionic conductivity leads to a more homogeneous concentration (blue $\to$ red $\to$ yellow), assigning the gap to the electrode leads to a lower concentration in the center and a higher concentration at the vertices of the surface of the electrode (dashed lines), and assigning the gap to the electrolyte leads to a higher concentration in the center and a lower concentration at the vertices of the surface of the electrode (dash-dotted lines). Furthermore, it can be seen that in typical setups, the deviation from the resolved model is dominated by the geometric change (dash-dotted lines), while the influence of the assumption of the normal component of the electric field being constant in the normal direction is rather negligible.
\subsection{Quantification of the influence of active material coating in a geometrically realistic microstructure}
\label{sec:influence_coating}
In this section, we show the applicability of the outlined approach to realistic microstructures, highlight the advantages of the outlined approach compared to other approaches, and gain new insights considering the impact of coating layers on the cell performance.
\subsubsection{Geometry}
\begin{figure}[ht]
    \def\svgwidth{0.4\textwidth}
    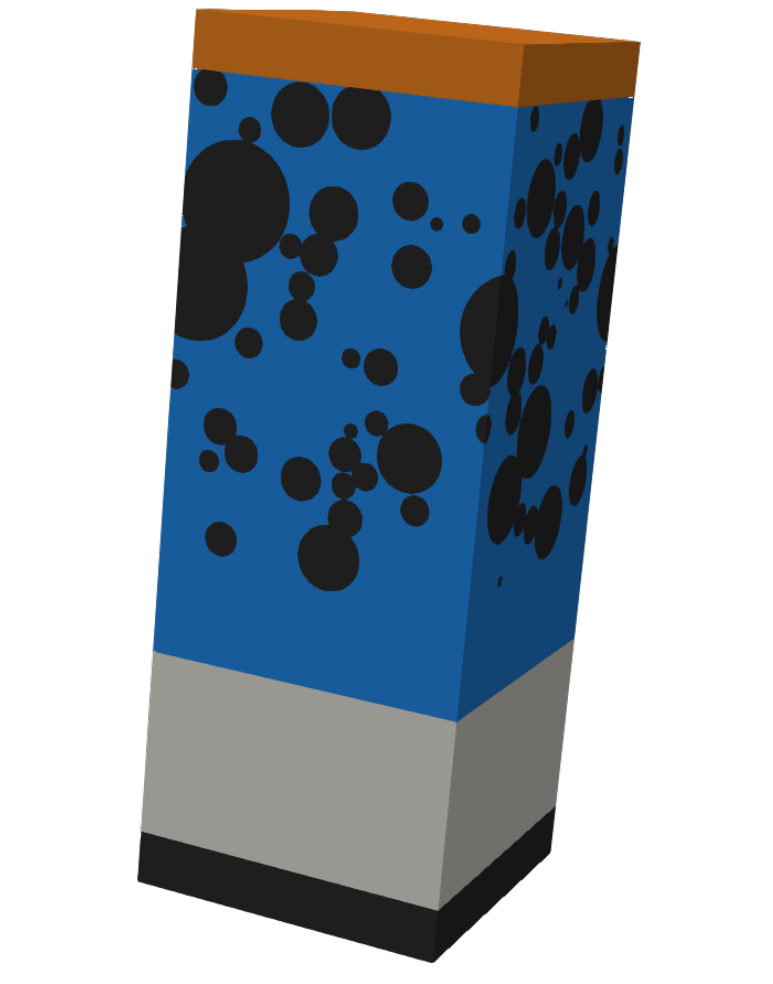
    \caption{Realistic microstructure including spherical cathode active material particles, a solid electrolyte, a metal anode, and the current collectors. Due to our proposed approach, the coating layer does not need to be spatially resolved and thus is not visible at the interface between cathode active material particles and the solid electrolyte.}
    \label{fig:realisitc_geometry}
\end{figure}

A representative geometric domain is required to gain quantitative insights into the influence of the active material coating on the cell performance. Therefore, we create the geometric domain of the composite cathode based on an artificial workflow, as similarly done in our previous work~\cite{Schmidt2022}, that satisfies the statistical properties of the composite electrode, i.e. the volume ratio of active material and solid electrolyte, the radius distribution of the as spheres approximated active material particles (e.g. NMC), and the porosity: Spheres are drawn from a log-normal distribution of the radius of the particles until the desired volumetric ratio is obtained. Afterwards, the spheres are spatially arranged using a simulation employing the discrete-element method until the steady state is reached. The solid electrolyte fills the voids between the particles in the composite cathode, i.e. the porosity is geometrically not resolved. The remaining components of the battery cell (the solid electrolyte separator, the metal anode, and the current collectors) are assumed as cuboid blocks. The obtained geometry is meshed with tetrahedral elements using \textit{Coreform Cubit 2021.3}, resulting in 564,711 nodes. The geometry resulting from this workflow is shown in \cref{fig:realisitc_geometry}, and the geometric parameters are summarized in \cref{table:parameters_realistic}.
\input{tables/parameters_realisitc.tex}
\subsubsection{Materials}
The material parameters are selected to match the setup in \cite{Neumann2020} and~\cite{Schmidt2022} and are extended by values for the coating layer. The cathode active material is a nickel manganese cobalt oxide (NMC622), the anode material is pure lithium metal, the electrolyte is a thiophosphate ($\beta$-LPS), the anode side current collector is copper, and the cathode side current collector is aluminum. For the coating material, lithium niobate ($\text{LiNbO}_3)$ is chosen, which is a common coating material for various active materials~\cite{Culver2019}. The material parameters are summarized in \cref{table:material_parameters_realistic}.
\subsubsection{Boundary and initial conditions}
A discharge scenario from the fully charged state with a constant external current is simulated. Therefore, the lower cut-off voltage is defined to $\Delta \Phi_\text{low} = \Phi_{\Gamma_\text{cc,c-o, low}} - \Phi_{\Gamma_\text{cc,a-o, low}}= 2.8 \ \text{V}$ and the C-rate is set to~$\hat{C}=0.5$, with an initial ramp to prevent oscillations ($C(t) = \hat{C} \frac{1}{2}(1+\text{cos}(\frac{\pi}{100 \ \text{s}} t)) \ \text{for} \ t < 100 \ \text{s}$ and $\hat{C}$ else). Of course, certain coating materials are used to enlarge the chemical stability window. However, we consider the lower cut-off voltage fixed to not mix up different effects. Initially, the lithium (-ion) concentrations in the domains are chosen to represent a fully charged state, i.e. $c_{0,\text{a}} = 7.69 \cdot 10^4 \frac{\text{mol}}{\text{m}^3}$, $c_{0,\text{c}} = 2.1 \cdot 10^4 \frac{\text{mol}}{\text{m}^3}$ (from $\text{SOC} = \frac{c_{0,\text{c}} - c_\text{max} \chi_{0\%}}{c_\text{max} (\chi_{100\%}-\chi_{0\%})} = 100 \%$), and $c_{0,\text{el}} = c_{0,\text{coat}} = 1.03 \cdot 10^4 \frac{\text{mol}}{\text{m}^3}$.
\subsubsection{Results of the simulations}
\begin{figure}[ht]
    \centering
    \input{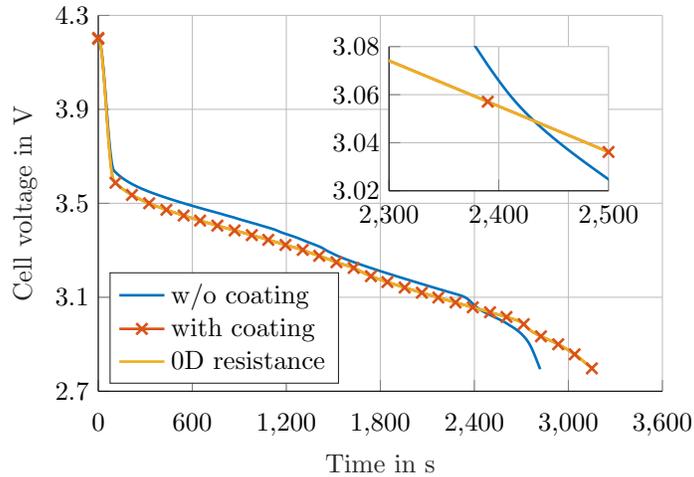}
    \caption{Comparison of the cell voltage of a battery cell including a coating layer ("with coating") and a battery cell without a coating layer ("w/o coating"). For comparison, the cell voltage computed with a model that incorporates the coating layer by a zero-dimensional resistance is added ("0D resistance").}
    \label{fig:realistic_cell_voltage_with_without_coating}
\end{figure}

\begin{figure}[H]
    \centering
    \input{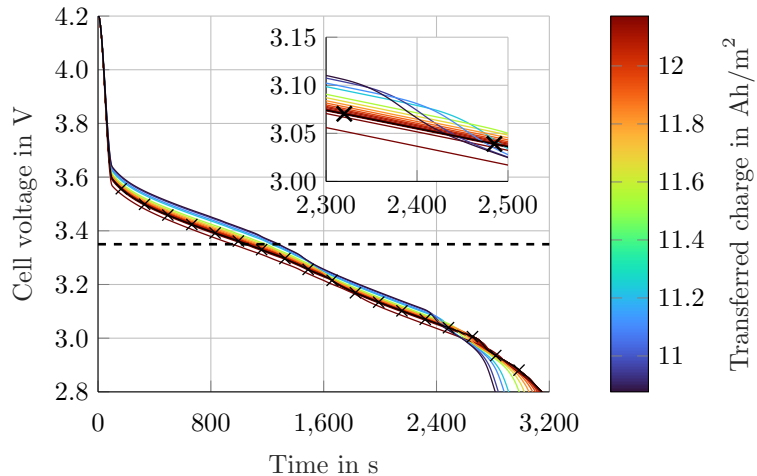}
    \caption{Cell voltage over time for the different parameter combinations. The color bar indicates the exchanged charge normalized by the lateral area of the battery cell. The black line with crosses represents the cell voltage for the default parameters.}
    \label{fig:realistic_parameters_cell_voltage}
\end{figure}

\begin{figure}[H]
    \centering
    \includegraphics{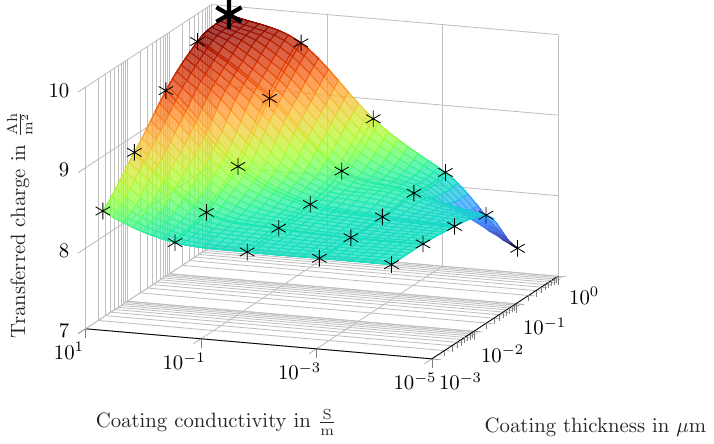}
    \caption{Dependence of transferred charge normalized by the lateral area on the ionic conductivity of the coating layer and on the thickness of the coating layer when the cell voltage has reached $\Delta \Phi = 3.35 \ \text{V}$. The dots denote the model evaluations. The bold dot denotes the setup with the optimal parameter combination.}
    \label{fig:realistic_influence_thickness_conductivity_middle}
\end{figure}

\begin{figure}[H]
    \centering
    \includegraphics{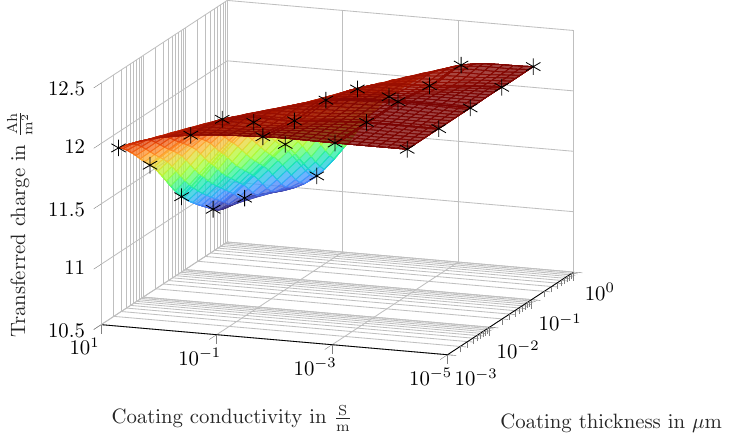}
    \caption{Dependence of transferred charge normalized by the lateral area on the ionic conductivity of the coating layer and on the thickness of the coating layer at the end of discharge. The dots denote the model evaluations.}
    \label{fig:realistic_influence_thickness_conductivity_end}
\end{figure}

\begin{figure}[ht]
    \centering
    \hfill
    \begin{subfigure}{0.4\textwidth}
        \centering
        \vfill
        \includegraphics[width = \textwidth]{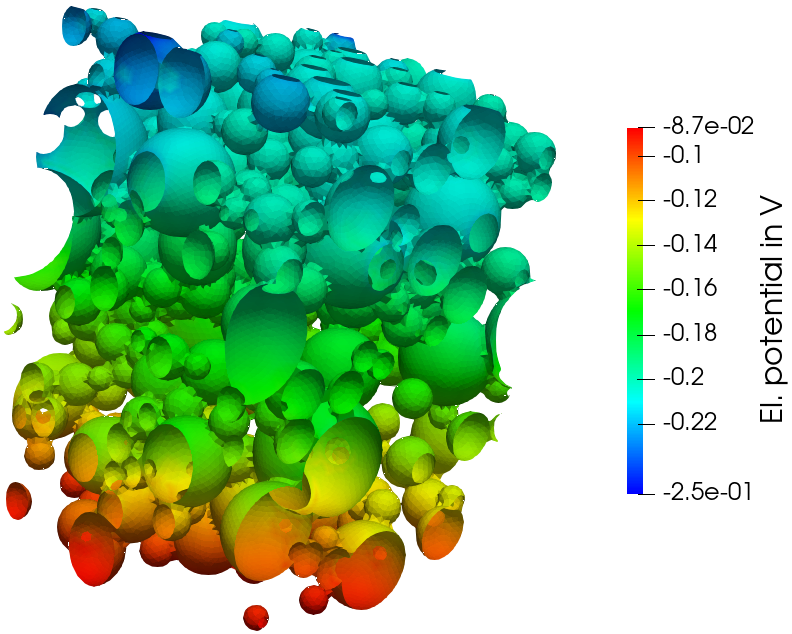}
        \caption{Electric potential in the coating layer with the default parameters at $t=1060 \ \text{s}$.}
        \label{fig:realistic_with_coating_el_pot_standard}
        \vfill
    \end{subfigure}
    \hfill
    \begin{subfigure}{0.4\textwidth}
        \centering
        \vfill
        \includegraphics[width = \textwidth]{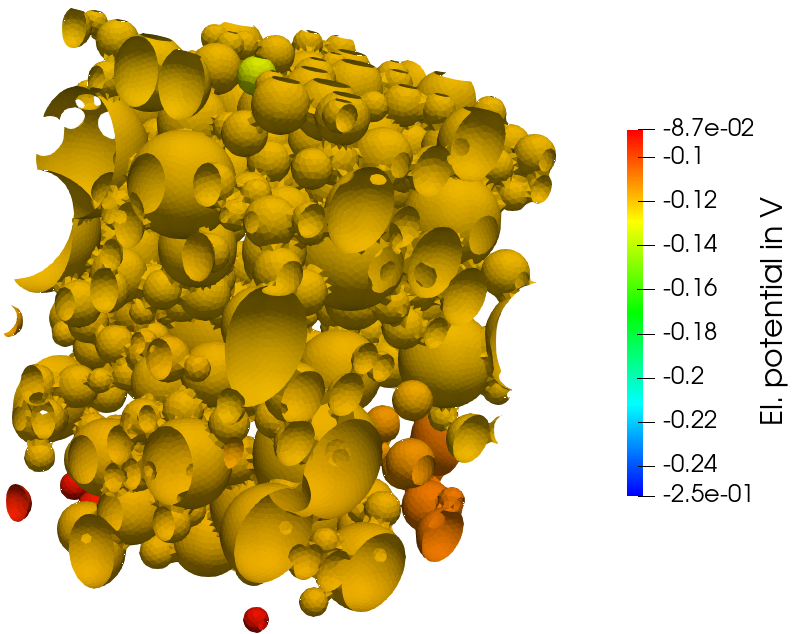}
        \caption{Electric potential in the coating layer with the optimal parameters at $t=1285 \ \text{s}$.}
        \label{fig:realistic_with_coating_el_pot_optimal}
        \vfill
    \end{subfigure}
    \hfill
    \caption{Electric potential in the coating layer when $\Delta \Phi = 3.35 \ \text{V}$ is reached. The current collector is attached at the top.}
    \label{fig:realistic_el_potential_coating_layer}
\end{figure}

\begin{figure}[H]
    \centering
    \hfill
    \begin{subfigure}{0.4\textwidth}
        \centering
        \vfill
        \includegraphics[width = \textwidth]{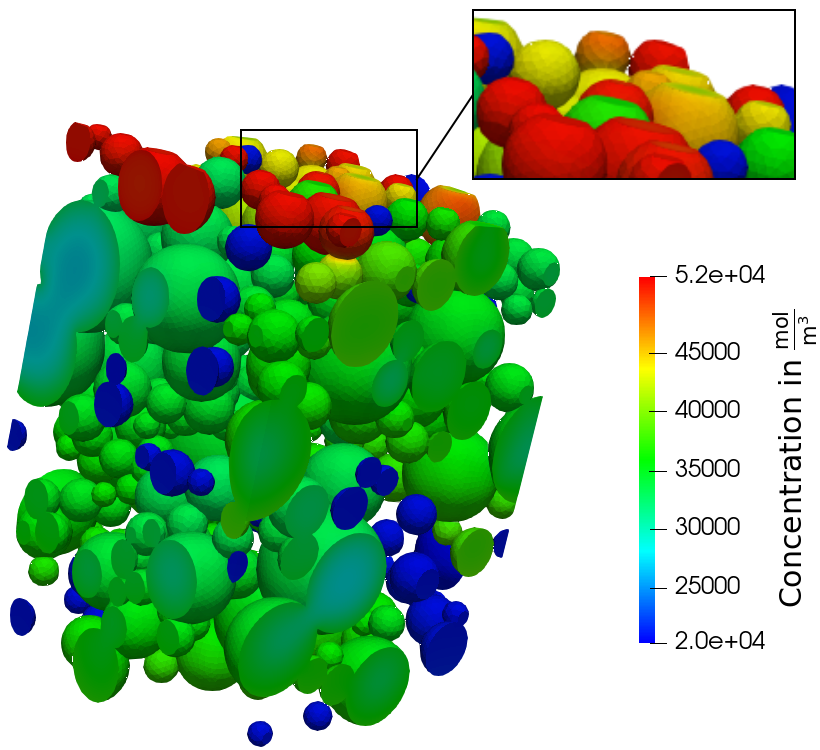}
        \caption{Concentration with default parameters at $t=3150 \ \text{s}$.}
        \label{fig:realistic_with_coating_conc_standard}
        \vfill
    \end{subfigure}
    \hfill
    \begin{subfigure}{0.4\textwidth}
        \centering
        \vfill
        \includegraphics[width = \textwidth]{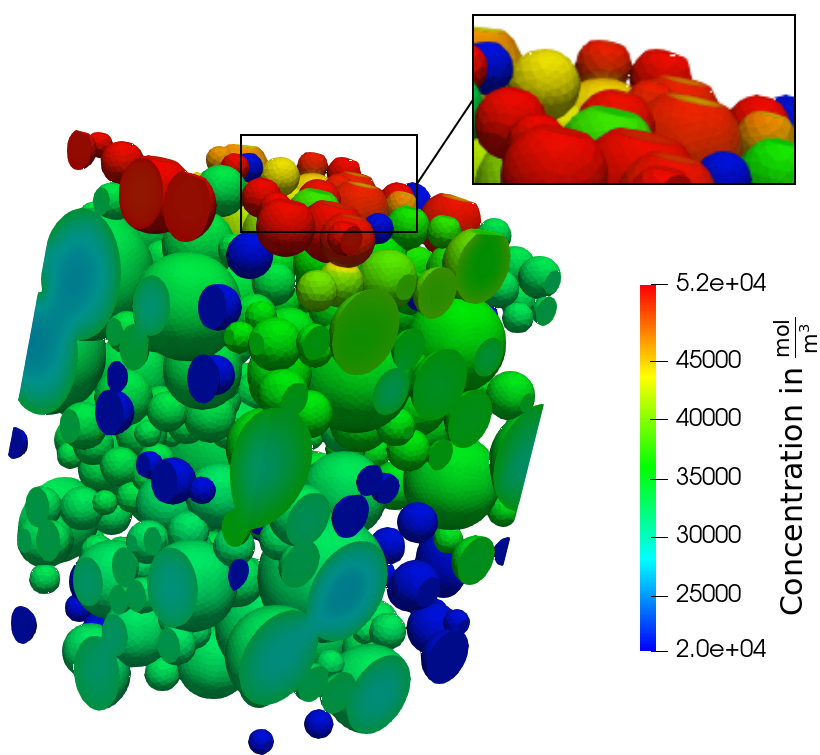}
        \caption{Concentration with optimal parameters at $t=2820 \ \text{s}$.}
        \label{fig:realistic_with_coating_conc_optimal}
        \vfill
    \end{subfigure}
    \hfill
    \caption{Concentration in the cathode at the end of discharge. The current collector is attached at the top.}
    \label{fig:realistic_concentration_cathode}
\end{figure}

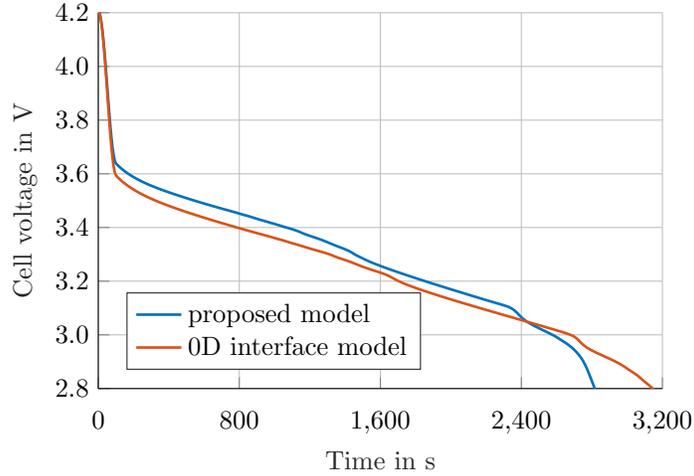
\begin{figure}[ht]
    \centering
    \input{figures/realistic_cell_voltage_optimal_without_resolved_coating.tikz}
    \caption{Comparison of the cell voltage with the proposed model and a model that only adds an interface resistance for the setup with optimal parameters.}
    \label{fig:realistic_cell_voltage_optimal_without_resolved_coating}
\end{figure}

At first, the results computed with the material parameters as summarized in \cref{table:material_parameters_realistic} (called "default parameters") are analyzed. Afterwards, we vary the most influential parameters of the coating layer to show the capabilities of the outlined approach compared to other approaches presented in the literature. We will restrict the discussion to the results of electrochemical quantities to limit the length of this paper. However, the mechanical state is solved as well.
\paragraph{Default parameters}
We compare the results computed with the material parameters as summarized in \cref{table:material_parameters_realistic} with the results computed without a coating layer and with the results computed with an existing approach that incorporates the coating layer by a simple zero-dimensional resistance. For the thickness of the coating layer, $t_\text{coat} = 10 \ \text{nm}$ is chosen as reported in the literature~\cite{Culver2019}. For the simulation without a coating layer, the domain $\Omega_\text{coat}$ is neglected, and thus, an interface between the cathode and the solid electrolyte~$\Gamma_\text{c-el}$ occurs where the kinetic laws of the interface between the cathode and the coating layer~$\Gamma_\text{coat-c}$ of the simulation including the coating layer are applied to. In \cref{fig:realistic_cell_voltage_with_without_coating}, the cell voltage computed for these different setups is compared.
As expected, the cell resistance slightly increases in the case of the coating layer due to the comparably poor conductivity of the coating layer, and thus, the cell voltage is lower until $t \approx 2400 \ \text{s}$ (the kink in the cell voltage at $t \approx 2400 \ \text{s}$ in the simulation without a coating layer is caused by the so-called 'sandwich-lithiation' which will be discussed later). The effect of the coating layer for this setup could, in a first-order approximation, be captured by a zero-dimensional resistance as the ionic conductivity of the coating layer is almost two orders of magnitude smaller compared to the ionic conductivity of the bulk electrolyte. Consequently, a short conduction path in the coating layer is favored, which means a negligible contribution to the conduction in the tangential direction. Therefore, the cell voltage computed with the model with a zero-dimensional resistance that cannot capture transport in the tangential direction of the coating layer almost perfectly matches the cell voltage computed with the outlined model.
\paragraph{Variation of the most influential coating parameters}
We systematically vary the most influential parameters of the coating layer to create setups where the optimal conduction path has significant contributions in the tangential direction of the coating layer to highlight the necessity of spatially resolving the coating layer compared to zero-dimensional interface models (e.g. in~\cite{Javed2020}). The most influential parameters of the coating layer can be identified without an elaborate sensitivity analysis as the ionic conductivity of the coating material and the thickness of the coating layer. These parameters~$v$ are systematically varied using a uniform on a logarithmic scale for both parameters with 5 samples each, i.e. 25 samples to cover all combinations. The bounds of the parameters are listed in \cref{table:material_parameters_variations}.
\input{tables/material_parameters_variations.tex}
We evaluate how the parameters influence the transferred charge. In \cref{fig:realistic_parameters_cell_voltage}, the development of the cell voltage over time for the 25 combinations is shown.
 \noindent
There exist parameter combinations that feature a higher cell voltage at the beginning and a lower cell voltage towards the end of the discharge compared to the cell voltage computed with the default parameters. This unexpected behavior is further investigated by plotting the parameters (ionic conductivity and thickness of the coating layer) together with the transferred charge normalized by the lateral area ($A=l_\text{l}^2 = 3600 \ \mu \text{m}^2$) in a three-dimensional parameter space after reaching a cell voltage of $\Delta \Phi = 3.35 \ \text{V}$ (see \cref{fig:realistic_influence_thickness_conductivity_middle}) and at the end of discharge at~$\Delta \Phi = 2.8 \ \text{V}$ (see \cref{fig:realistic_influence_thickness_conductivity_end}).
 \noindent
While at the beginning of the discharge (while $\Delta \Phi > 3.35 \ \text{V}$), a larger thickness of the coating layer together with a high ionic conductivity is advantageous, the opposite holds at the end of charging (while $2.8 \ \text{V} < \Delta \Phi < 3.35 \ \text{V}$). Both observations can be assigned to different phenomena that occur concurrently but dominate at different times during the discharge. We will first discuss the development of the cell voltage until $\Delta \Phi = 3.35 \ \text{V}$ is reached and afterwards the development of the cell voltage until the end of discharge.\\
The preferred parameter combination in \cref{fig:realistic_influence_thickness_conductivity_middle} can be explained by improved conduction paths that result from the good conductivity and the thick coating layer. This improved conduction in the coating layer is expressed in terms of the gradient of the electric potential in the coating layer. In \cref{fig:realistic_el_potential_coating_layer}, the electric potential in the coating layer is compared for the case where the most charge is transferred until $\Delta \Phi = 3.35 \ \text{V}$ ($\kappa=5 \frac{\text{S}}{\text{m}}$, $t_\text{coat}=1 \ \mu \text{m}$) is reached (labeled with "optimal parameters") with the results from the default set of parameters (labeled with "default parameters"), i.e. at $t=1060 \ \text{s}$ for the default parameters and at $t=1285 \ \text{s}$ for the optimal parameters.
Obviously, the magnitude of the gradient is smaller in the setup with the optimal parameters, and thus, the total resistance of the cell reduces, and more charge is transferred until $\Delta \Phi = 3.35 \ \text{V}$ is reached. \\
The averaged current in the $z$-direction in the coating layer is~$\bar{i}_z = \frac{\int_{\Gamma_{\text{coat}}} - \kappa_\text{coat} \nabla \Phi \cdot \vec{e}_z \ \text{d}\Gamma}{\int_{\Gamma_{\text{coat}}} \ \text{d}\Gamma}$ with the unit vector $\vec{e}_z$ in $z$-direction. It serves as an indicator of how much charge is transported along the coating layer. For the default parameters, the current in z-direction is~$\bar{i}_{z,\text{def}}= -0.515 \frac{\text{A}}{\text{m}^2}$, and for the optimal parameters it is~$\bar{i}_{z,\text{opt}}= -21.3 \frac{\text{A}}{\text{m}^2}$. This shows that more charge is transported along the coating layer for the setup with the optimal parameters compared to the setup with the default parameters due to the favored conduction paths in the coating layer. \\ 
However, the parameter combinations in \cref{fig:realistic_influence_thickness_conductivity_end} show that a good conductivity and a large thickness of the coating layer are disadvantageous if the transferred charge until $\Delta \Phi = 2.8 \ \text{V}$ is considered. This can be explained by the so-called 'sandwich lithiation' that is observable for some material combinations \cite{Neumann2020}. Due to the dependence of the electronic conductivity in the cathode active material on the degree of lithiation, the electronic conductivity is lowered towards a higher degree of lithiation and, thus, becomes the limiting factor towards the end of discharge. Increasing the ionic conductivity of the coating layer stresses this limitation as the ratio of the ionic conductivity in the coating layer and the electronic conductivity in the active material increases. By comparing the three-dimensionally resolved concentrations at the end of discharge in \cref{fig:realistic_concentration_cathode} for the setup with optimal parameters (at $t = 2820 \ \text{s}$), and the setup with default parameters (at $t = 3150 \ \text{s}$), the different values for the concentration close to the current collector become visible: the delithiation in the setup with optimal parameters is more inhomogeneous towards the end of discharge. Due to the smaller electronic conductivity originating from its lithiation dependence, the lower cut-off voltage is reached earlier, and thus, less charge is transferred.
\paragraph{Neglecting conduction along the coating layer}
Finally, we compare the results for the setup with the optimal parameters with a model where the coating layer is modeled using a zero-dimensional resistance, i.e. only the normal flux density in the coating layer is considered while the tangential contribution is neglected. In that setup, the interface resistance is set to $\bar{r}_\text{i} = r_\text{i} + \left.\frac{t_\text{coat}}{\kappa}\right|_{\text{opt}} = 5.0002 \cdot 10^{-3} \ \Omega \text{m}^2$. The cell voltage computed with both models is shown in \cref{fig:realistic_cell_voltage_optimal_without_resolved_coating}.
It is evident that the results of both models significantly deviate, as only the outlined approach considers conduction paths along the coating layer, while the other model only adds a resistance to the interface. The previously outlined 'sandwich lithiation' is less pronounced with a zero-dimensional interface law. This highlights that representing the coating layer with a zero-dimensional interface model is in general not sufficient. We can conclude that conduction along the coating layer becomes significant if the conductivity of the coating layer reaches the magnitude of the conductivity of the bulk electrolyte, and subsequently, an approach that considers this tangential conduction becomes mandatory.

%% file: figures/geometryRepresentationSphere.pdf_tex
%% Creator: Inkscape inkscape 0.92.5, www.inkscape.org
%% Accompanies image file 'geometryRepresentationSphere.pdf' (pdf, eps, ps)
\begingroup%
  \makeatletter%
  \providecommand\color[2][]{%
    \errmessage{(Inkscape) Color is used for the text in Inkscape, but the package 'color.sty' is not loaded}%
    \renewcommand\color[2][]{}%
  }%
  \providecommand\transparent[1]{%
    \errmessage{(Inkscape) Transparency is used (non-zero) for the text in Inkscape, but the package 'transparent.sty' is not loaded}%
    \renewcommand\transparent[1]{}%
  }%
  \providecommand\rotatebox[2]{#2}%
  \newcommand*\fsize{\dimexpr\f@size pt\relax}%
  \newcommand*\lineheight[1]{\fontsize{\fsize}{#1\fsize}\selectfont}%
  \ifx\svgwidth\undefined%
    \setlength{\unitlength}{398.0498512bp}%
    \ifx\svgscale\undefined%
      \relax%
    \else%
      \setlength{\unitlength}{\unitlength * \real{\svgscale}}%
    \fi%
  \else%
    \setlength{\unitlength}{\svgwidth}%
  \fi%
  \global\let\svgwidth\undefined%
  \global\let\svgscale\undefined%
  \makeatother%
  \begin{picture}(1,0.89403768)%
    \lineheight{1}%
    \setlength\tabcolsep{0pt}%
    \put(0,0){\includegraphics[width=\unitlength,page=1]{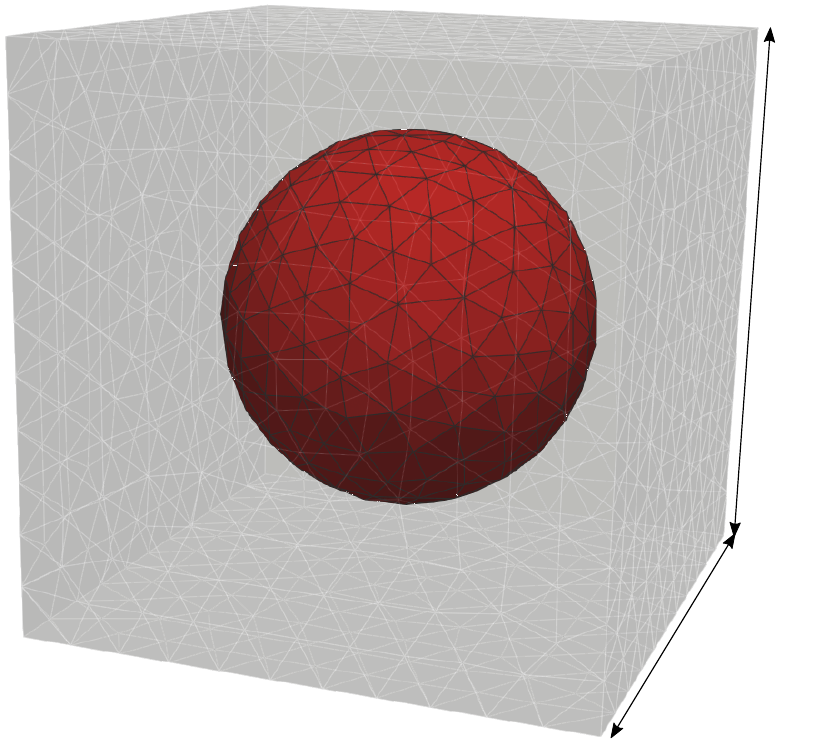}}%
    \put(0.92755697,0.5382641){\color[rgb]{0,0,0}\makebox(0,0)[lt]{\lineheight{1.25}\smash{\begin{tabular}[t]{l}$a$\end{tabular}}}}%
    \put(0.83626238,0.12119604){\color[rgb]{0,0,0}\makebox(0,0)[lt]{\lineheight{1.25}\smash{\begin{tabular}[t]{l}$a$\end{tabular}}}}%
    \put(0,0){\includegraphics[width=\unitlength,page=2]{geometryRepresentationSphere.pdf}}%
    \put(0.49707783,0.74535247){\color[rgb]{0,0,0}\makebox(0,0)[lt]{\lineheight{1.25}\smash{\begin{tabular}[t]{l}$D$\end{tabular}}}}%
  \end{picture}%
\endgroup%

%% file: tables/test_setup_sphere.tex
\begin{tabular}{c | c | c}
    \hline
    \textbf{quantity}       & \textbf{symbol}          & \textbf{value}                          \\
    \hline
    side length of cube     &$a$          & $3 \ \text{mm}$                   \\
    diameter of sphere        & $D$     & $2 \ \text{mm}$                               \\
    thickness of coating & $t_\text{coat}$ & $150 \ \text{nm}$  \\
    \hline
\end{tabular}

%% file: figures/temporal_convergence_concentration_sphere.tikz
\definecolor{mycolor1}{rgb}{0.00000,0.44700,0.74100}%
\definecolor{mycolor2}{rgb}{0.85000,0.32500,0.09800}%
\definecolor{mycolor3}{rgb}{0.92900,0.69400,0.12500}%
\begin{tikzpicture}

\begin{axis}[%
width=7.5cm,
height=5.0cm,
scale only axis,
xmode=log,
xmin=0.1,
xmax=100,
xminorticks=true,
xlabel style={font=\color{white!15!black}},
xlabel={Size of time step in \text{s}},
x tick label style={
/pgf/number format/.cd,
/tikz/.cd,
yshift=-.5em},
ymode=log,
ymin=1.0e-7,
ymax=1.0,
y tick label style={
/pgf/number format/.cd,
/tikz/.cd},
yminorticks=true,
ylabel style={font=\color{white!15!black}},
ylabel={Maximal value of relative L2-norm $\epsilon$},
axis background/.style={fill=white},
xmajorgrids,
xminorgrids,
ymajorgrids,
yminorgrids
]
\addplot [color=mycolor1, line width=1.0pt, mark size=4.0pt, mark=asterisk, mark options={solid, mycolor1}, forget plot]
  table[row sep=crcr]{%
0.1	8.20063103267664e-07\\
1	7.49826431276635e-05\\
10	0.0075187683105471\\
100	1\\
};
\addplot [color=mycolor2, dashed, line width=1.0pt, forget plot]
  table[row sep=crcr]{%
0.1	1e-06\\
1	0.0001\\
10	0.01\\
100	1\\
};
\addplot [color=mycolor3, dashed, line width=1.0pt, forget plot]
  table[row sep=crcr]{%
0.1	0.001\\
1	0.01\\
10	0.1\\
100	1\\
};
\end{axis}
\end{tikzpicture}%

%% file: figures/comparison_geometry_squared.pdf_tex
%% Creator: Inkscape inkscape 0.92.5, www.inkscape.org
%% Accompanies image file 'comparison_geometry_squared.pdf' (pdf, eps, ps)
\begingroup%
  \makeatletter%
  \providecommand\color[2][]{%
    \errmessage{(Inkscape) Color is used for the text in Inkscape, but the package 'color.sty' is not loaded}%
    \renewcommand\color[2][]{}%
  }%
  \providecommand\transparent[1]{%
    \errmessage{(Inkscape) Transparency is used (non-zero) for the text in Inkscape, but the package 'transparent.sty' is not loaded}%
    \renewcommand\transparent[1]{}%
  }%
  \providecommand\rotatebox[2]{#2}%
  \newcommand*\fsize{\dimexpr\f@size pt\relax}%
  \newcommand*\lineheight[1]{\fontsize{\fsize}{#1\fsize}\selectfont}%
  \ifx\svgwidth\undefined%
    \setlength{\unitlength}{522.29459767bp}%
    \ifx\svgscale\undefined%
      \relax%
    \else%
      \setlength{\unitlength}{\unitlength * \real{\svgscale}}%
    \fi%
  \else%
    \setlength{\unitlength}{\svgwidth}%
  \fi%
  \global\let\svgwidth\undefined%
  \global\let\svgscale\undefined%
  \makeatother%
  \begin{picture}(1,0.92133699)%
    \lineheight{1}%
    \setlength\tabcolsep{0pt}%
    \put(0,0){\includegraphics[width=\unitlength,page=1]{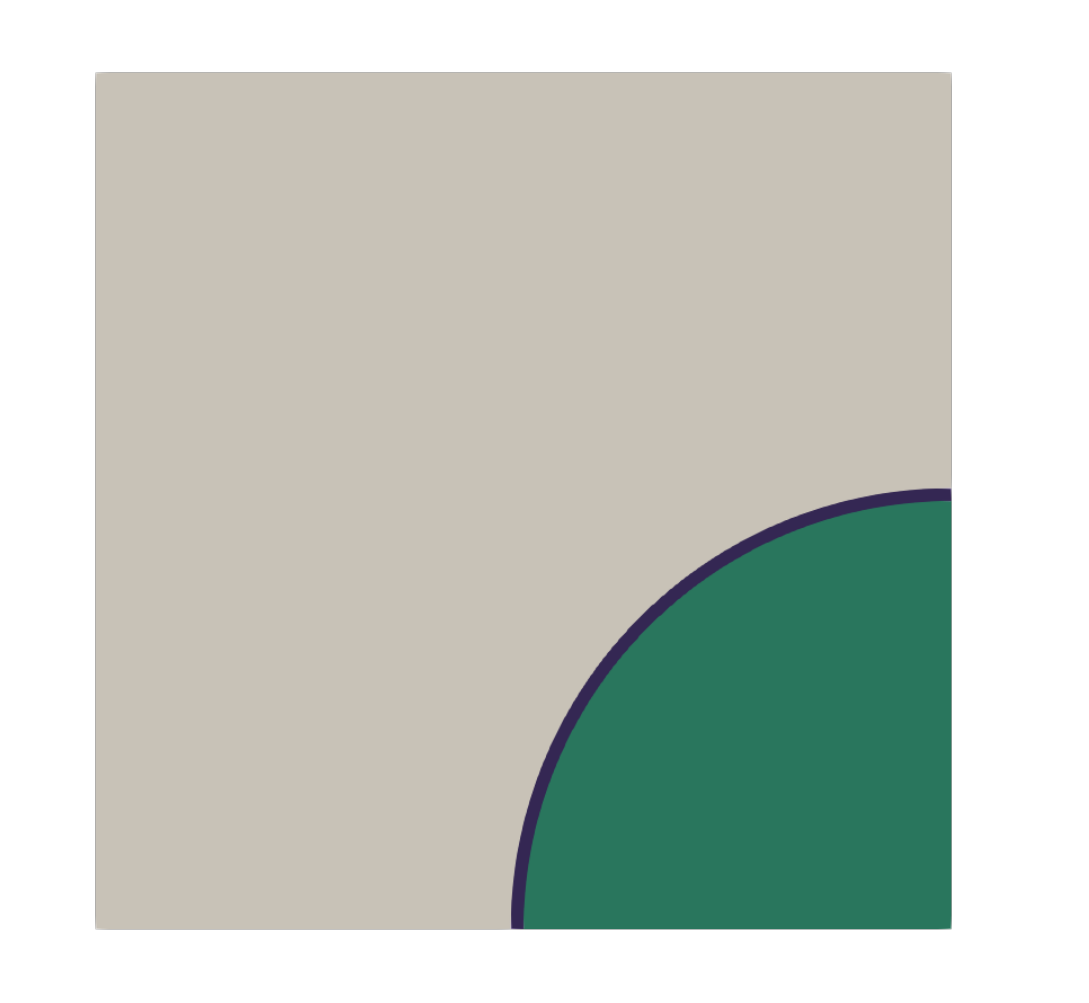}}%
    \put(0.9095701,0.26324598){\color[rgb]{0,0,0}\makebox(0,0)[lt]{\lineheight{1.25}\smash{\begin{tabular}[t]{l}$r$\end{tabular}}}}%
    \put(0,0){\includegraphics[width=\unitlength,page=2]{comparison_geometry_squared.pdf}}%
    \put(0.89597623,0.46934137){\color[rgb]{0,0,0}\makebox(0,0)[lt]{\lineheight{1.25}\smash{\begin{tabular}[t]{l}coating\\layer\end{tabular}}}}%
    \put(0.41088453,0.63447692){\color[rgb]{0,0,0}\makebox(0,0)[lt]{\lineheight{1.25}\smash{\begin{tabular}[t]{l}solid electrolyte\end{tabular}}}}%
    \put(0.63396411,0.21001456){\color[rgb]{0,0,0}\makebox(0,0)[lt]{\lineheight{1.25}\smash{\begin{tabular}[t]{l}\color{white}electrode\end{tabular}}}}%
    \put(0,0){\includegraphics[width=\unitlength,page=3]{comparison_geometry_squared.pdf}}%
    \put(0.45795935,0.89224362){\color[rgb]{0,0,0}\makebox(0,0)[lt]{\lineheight{1.25}\smash{\begin{tabular}[t]{l}$a$\end{tabular}}}}%
    \put(0,0){\includegraphics[width=\unitlength,page=4]{comparison_geometry_squared.pdf}}%
    \put(-0.00317858,0.44753308){\color[rgb]{0,0,0}\makebox(0,0)[lt]{\lineheight{1.25}\smash{\begin{tabular}[t]{l}$a$\end{tabular}}}}%
    \put(0,0){\includegraphics[width=\unitlength,page=5]{comparison_geometry_squared.pdf}}%
    \put(0.25801322,0.71137294){\color[rgb]{0,0,0}\makebox(0,0)[lt]{\lineheight{1.25}\smash{\begin{tabular}[t]{l}$l_\text{c}$\end{tabular}}}}%
    \put(0.23023918,0.08845483){\color[rgb]{0,0,0}\makebox(0,0)[lt]{\lineheight{1.25}\smash{\begin{tabular}[t]{l}$l_\text{v}$\end{tabular}}}}%
    \put(0,0){\includegraphics[width=\unitlength,page=6]{comparison_geometry_squared.pdf}}%
    \put(0.4567105,0.00180618){\color[rgb]{0,0,0}\makebox(0,0)[lt]{\lineheight{1.25}\smash{\begin{tabular}[t]{l}$t_\text{coat}$\end{tabular}}}}%
    \put(0,0){\includegraphics[width=\unitlength,page=7]{comparison_geometry_squared.pdf}}%
    \put(0.45402182,0.25862469){\color[rgb]{0,0,0}\makebox(0,0)[lt]{\lineheight{1.25}\smash{\begin{tabular}[t]{l}$t_1$\end{tabular}}}}%
    \put(0.36498284,0.09635311){\color[rgb]{0,0,0}\makebox(0,0)[lt]{\lineheight{1.25}\smash{\begin{tabular}[t]{l}$n$\end{tabular}}}}%
    \put(0,0){\includegraphics[width=\unitlength,page=8]{comparison_geometry_squared.pdf}}%
    \put(0.75919813,0.07633146){\color[rgb]{0,0,0}\makebox(0,0)[lt]{\lineheight{1.25}\smash{\begin{tabular}[t]{l}$\alpha$\end{tabular}}}}%
    \put(0,0){\includegraphics[width=\unitlength,page=9]{comparison_geometry_squared.pdf}}%
  \end{picture}%
\endgroup%

%% file: tables/parameters_comparison_resolved.tex
\begin{tabular}{c | c | c}
    \hline
    \textbf{quantity}                & \textbf{symbol}          & \textbf{value} \\
    \hline
    side length            & $a$ & $10 \ \mu \text{m}$\\
    radius of electrode    & $r$ & $5 \ \mu \text{m}$ \\
    thickness of coating & $t_\text{coat}$ & $150 \ \text{nm}$  \\
    \hline
\end{tabular}

%% file: figures/squared_geometry_standard_cond_radial_gradient_plot.tikz
\definecolor{mycolor1}{rgb}{0.00000,0.44700,0.74100}%
\definecolor{mycolor2}{rgb}{0.85000,0.32500,0.09800}%
\definecolor{mycolor3}{rgb}{0.92900,0.69400,0.12500}%
\begin{tikzpicture}

\begin{axis}[%
width=7.5cm,
height=5.0cm,
scale only axis,
xmin=0.0,
xmax=0.3,
xtick={0, 0.1, 0.2, 0.3},
xlabel style={font=\color{white!15!black}},
xlabel={Position in normal direction in $\mu\text{m}$},
ymin=-5.05270433312207,
ymax=5.05258035546096,
 tick label style={
/pgf/number format/.cd,
fixed,
fixed zerofill,
precision=1,
/tikz/.cd,
yshift=-.5em},
ymin=-12.0,
ymax=0,
ytick ={0, -3, -6, -9, -12},
y tick label style={
/pgf/number format/.cd,
fixed,
fixed zerofill,
precision=1,
/tikz/.cd},
ylabel style={font=\color{white!15!black}},
ylabel={Normal comp. of el. field in $\frac{\text{V}}{\text{mm}}$},
axis background/.style={fill=white},
axis x line*=bottom,
axis y line*=left,
xmajorgrids,
ymajorgrids,
legend style={at={(0.05,0.15)}, anchor=south west, legend cell align=left, align=left, draw=white!15!black}
]

\addplot [color=mycolor1, line width=1.0pt, forget plot]
  table[row sep=crcr]{%
0	-5.05258035773484\\
0.015	-5.04478836125531\\
0.03	-5.03699636477579\\
0.045	-5.02148961944891\\
0.06	-5.00598287321865\\
0.075	-4.99064731336732\\
0.09	-4.97531175723535\\
0.105	-4.96015405029158\\
0.12	-4.94499635439708\\
0.135	-4.93745994310711\\
0.15	-4.92992353547278\\
};

\addplot [color=mycolor2, dashed, line width=1.0pt, forget plot]
  table[row sep=crcr]{%
9.4e-05	-5.10164074024068\\
0.030085	-5.08687440858899\\
0.060076	-5.05784441462933\\
0.090067	-5.02917793516084\\
0.120058	-5.00086058627054\\
0.15005	-4.97288345939696\\
0.180041	-4.9452252024698\\
0.210032	-4.91790152758969\\
0.240023	-4.89090915942966\\
0.270014	-4.864329908996\\
0.300006	-4.83775065856234\\
};

\addplot [color=mycolor3, line width=1.0pt, forget plot]
  table[row sep=crcr]{%
0	-10.1250921809998\\
0.015	-10.0953053172832\\
0.03	-10.0655184535666\\
0.045	-10.0066433522944\\
0.06	-9.9477682510221\\
0.075	-9.89025921340378\\
0.09	-9.83275017578545\\
0.105	-9.77656050311291\\
0.12	-9.72037083044037\\
0.135	-9.69260162065569\\
0.15	-9.66483241087101\\
};
\end{axis}
\end{tikzpicture}%

%% file: figures/deviation_conduction_path.pdf_tex
%% Creator: Inkscape inkscape 0.92.5, www.inkscape.org
%% Accompanies image file 'deviation_conduction_path.pdf' (pdf, eps, ps)
\begingroup%
  \makeatletter%
  \providecommand\color[2][]{%
    \errmessage{(Inkscape) Color is used for the text in Inkscape, but the package 'color.sty' is not loaded}%
    \renewcommand\color[2][]{}%
  }%
  \providecommand\transparent[1]{%
    \errmessage{(Inkscape) Transparency is used (non-zero) for the text in Inkscape, but the package 'transparent.sty' is not loaded}%
    \renewcommand\transparent[1]{}%
  }%
  \providecommand\rotatebox[2]{#2}%
  \newcommand*\fsize{\dimexpr\f@size pt\relax}%
  \newcommand*\lineheight[1]{\fontsize{\fsize}{#1\fsize}\selectfont}%
  \ifx\svgwidth\undefined%
    \setlength{\unitlength}{94.43743077bp}%
    \ifx\svgscale\undefined%
      \relax%
    \else%
      \setlength{\unitlength}{\unitlength * \real{\svgscale}}%
    \fi%
  \else%
    \setlength{\unitlength}{\svgwidth}%
  \fi%
  \global\let\svgwidth\undefined%
  \global\let\svgscale\undefined%
  \makeatother%
  \begin{picture}(1,0.7049149)%
    \lineheight{1}%
    \setlength\tabcolsep{0pt}%
    \put(0,0){\includegraphics[width=\unitlength,page=1]{deviation_conduction_path.pdf}}%
    \put(0.43131706,0.20346052){\color[rgb]{0,0,0}\makebox(0,0)[lt]{\lineheight{1.25}\smash{\begin{tabular}[t]{l}$A$\end{tabular}}}}%
    \put(0.06738346,0.52334732){\color[rgb]{0,0,0}\makebox(0,0)[lt]{\lineheight{1.25}\smash{\begin{tabular}[t]{l}$B$\end{tabular}}}}%
    \put(0.27571719,0.46814891){\color[rgb]{0,0,0}\makebox(0,0)[lt]{\lineheight{1.25}\smash{\begin{tabular}[t]{l}$\Omega_\text{coat}$\end{tabular}}}}%
    \put(-0.00527383,0.20938716){\color[rgb]{0,0,0}\makebox(0,0)[lt]{\lineheight{1.25}\smash{\begin{tabular}[t]{l}$\Omega_\text{el}$\end{tabular}}}}%
  \end{picture}%
\endgroup%

%% file: figures/realisitc_geometry.pdf_tex
%% Creator: Inkscape inkscape 0.92.5, www.inkscape.org
%% Accompanies image file 'realisitc_geometry.pdf' (pdf, eps, ps)
\begingroup%
  \makeatletter%
  \providecommand\color[2][]{%
    \errmessage{(Inkscape) Color is used for the text in Inkscape, but the package 'color.sty' is not loaded}%
    \renewcommand\color[2][]{}%
  }%
  \providecommand\transparent[1]{%
    \errmessage{(Inkscape) Transparency is used (non-zero) for the text in Inkscape, but the package 'transparent.sty' is not loaded}%
    \renewcommand\transparent[1]{}%
  }%
  \providecommand\rotatebox[2]{#2}%
  \newcommand*\fsize{\dimexpr\f@size pt\relax}%
  \newcommand*\lineheight[1]{\fontsize{\fsize}{#1\fsize}\selectfont}%
  \ifx\svgwidth\undefined%
    \setlength{\unitlength}{370.72825756bp}%
    \ifx\svgscale\undefined%
      \relax%
    \else%
      \setlength{\unitlength}{\unitlength * \real{\svgscale}}%
    \fi%
  \else%
    \setlength{\unitlength}{\svgwidth}%
  \fi%
  \global\let\svgwidth\undefined%
  \global\let\svgscale\undefined%
  \makeatother%
  \begin{picture}(1,1.2737944)%
    \lineheight{1}%
    \setlength\tabcolsep{0pt}%
    \put(0,0){\includegraphics[width=\unitlength,page=1]{realisitc_geometry.pdf}}%
    \put(0.31357419,0.27226713){\color[rgb]{0,0,0}\rotatebox{-13}{\makebox(0,0)[lt]{\lineheight{1.25}\smash{\begin{tabular}[t]{l}anode\end{tabular}}}}}%
    \put(0.31426849,0.45140726){\color[rgb]{0,0,0}\rotatebox{-13}{\makebox(0,0)[lt]{\lineheight{1.25}\smash{\begin{tabular}[t]{l}separator\end{tabular}}}}}%
    \put(-0.00074415,0.92901473){\color[rgb]{0,0,0}\makebox(0,0)[lt]{\lineheight{1.25}\smash{\begin{tabular}[t]{l}composite \\\end{tabular}}}}%
    \put(0.28415952,1.20704573){\color[rgb]{0,0,0}\rotatebox{-6}{\makebox(0,0)[lt]{\lineheight{1.25}\smash{\begin{tabular}[t]{l}current collector\end{tabular}}}}}%
    \put(0.18317882,0.15411576){\color[rgb]{0,0,0}\rotatebox{-15}{\makebox(0,0)[lt]{\lineheight{1.25}\smash{\begin{tabular}[t]{l}\color{white}current collector\end{tabular}}}}}%
    \put(0,0){\includegraphics[width=\unitlength,page=2]{realisitc_geometry.pdf}}%
    \put(0.85712243,1.17005246){\color[rgb]{0,0,0}\makebox(0,0)[lt]{\lineheight{1.25}\smash{\begin{tabular}[t]{l}$l_\text{cc}$\end{tabular}}}}%
    \put(0.82408613,0.86276673){\color[rgb]{0,0,0}\makebox(0,0)[lt]{\lineheight{1.25}\smash{\begin{tabular}[t]{l}$l_\text{c}$\end{tabular}}}}%
    \put(0.78301118,0.49327891){\color[rgb]{0,0,0}\makebox(0,0)[lt]{\lineheight{1.25}\smash{\begin{tabular}[t]{l}$l_\text{se}$\end{tabular}}}}%
    \put(0.7618331,0.31200572){\color[rgb]{0,0,0}\makebox(0,0)[lt]{\lineheight{1.25}\smash{\begin{tabular}[t]{l}$l_\text{a}$\end{tabular}}}}%
    \put(0.75836927,0.19654508){\color[rgb]{0,0,0}\makebox(0,0)[lt]{\lineheight{1.25}\smash{\begin{tabular}[t]{l}$l_\text{cc}$\end{tabular}}}}%
    \put(0,0){\includegraphics[width=\unitlength,page=3]{realisitc_geometry.pdf}}%
    \put(0.31126941,0.02265173){\color[rgb]{0,0,0}\makebox(0,0)[lt]{\lineheight{1.25}\smash{\begin{tabular}[t]{l}$l_\text{l}$\end{tabular}}}}%
    \put(0,0){\includegraphics[width=\unitlength,page=4]{realisitc_geometry.pdf}}%
    \put(0.89664497,0.7240801){\color[rgb]{0,0,0}\makebox(0,0)[lt]{\lineheight{1.25}\smash{\begin{tabular}[t]{l}$z$\end{tabular}}}}%
    \put(0.01911319,0.87205164){\color[rgb]{0,0,0}\makebox(0,0)[lt]{\lineheight{1.25}\smash{\begin{tabular}[t]{l} cathode\end{tabular}}}}%
  \end{picture}%
\endgroup%

%% file: tables/parameters_realisitc.tex
\begin{table}[ht]
    \renewcommand{\arraystretch}{1.1}
    \centering
    \caption{Geometric parameters of the realistic microstructure.}
    \label{table:parameters_realistic}
    \begin{tabular}{c | c | c}
        \hline
        \textbf{quantity}              & \textbf{symbol}               & \textbf{value} \\
        \hline
        length of current collectors   & $l_\text{cc}$                                                & $10 \ \mu \text{m}$ \\
        length of composite cathode    & $l_\text{c}$                                       & $85 \ \mu \text{m}$ \\
        length of separator            & $l_\text{se}$                                       & $20 \ \mu \text{m}$ \\
        length of anode                & $l_\text{a}$                                       & $35 \ \mu \text{m}$ \\
        lateral length                 & $l_\text{l}$                                       & $60 \ \mu \text{m}$ \\
        log-normal distribution of diameter of cathode particles                       & \begin{tabular}{c} $\mu$ \\ $\sigma$ \end{tabular} & \begin{tabular}{c} $1.8189$ \\ $0.4589$ \end{tabular} \\
        volumetric ratio of AM and SE in composite cathode                             & $r$                                                & $0.499$ \\
        \hline
    \end{tabular}
\end{table}

%% file: tables/material_parameters_variations.tex
\begin{table}[ht]
    \renewcommand{\arraystretch}{1.1}
    \centering
    \caption{Range of parameters used for the variation.}
    \begin{tabular}{c | c | c}
      \hline
      \textbf{quantity}             & \textbf{min. value}   & \textbf{max. value}      \\
      \hline
      ionic conductivity of coating & $5 \cdot 10^{-5} \ \frac{\text{S}}{\text{m}}$ & $5 \ \frac{\text{S}}{\text{m}}$ \\
      thickness of coating & $10^{-3} \mu \text{m}$ & $1 \ \mu \text{m}$ \\
      \hline
    \end{tabular}
    \label{table:material_parameters_variations}
\end{table}

%% file: figures/realistic_cell_voltage_optimal_without_resolved_coating.tikz
\definecolor{mycolor1}{rgb}{0.00000,0.44700,0.74100}%
\definecolor{mycolor2}{rgb}{0.85000,0.32500,0.09800}%
\begin{tikzpicture}

\begin{axis}[%
width=7.5cm,
height=5.0cm,
scale only axis,
xmin=0,
xmax=3200,
xtick={0,800,1600,2400,3200},
xlabel style={font=\color{white!15!black}},
xlabel={Time in s},
x tick label style={
/pgf/number format/.cd,
fixed,
fixed zerofill,
precision=0,
/tikz/.cd,
yshift=-.5em},
ymin=2.8,
ymax=4.2,
ytick ={2.8,3.0,3.2,3.4,3.6,3.8,4.0,4.2},
y tick label style={
/pgf/number format/.cd,
fixed,
fixed zerofill,
precision=1,
/tikz/.cd},
ylabel style={font=\color{white!15!black}},
ylabel={Cell voltage in V},
axis background/.style={fill=white},
axis x line*=bottom,
axis y line*=left,
xmajorgrids,
ymajorgrids,
legend style={at={(0.05,0.05)}, anchor=south west, legend cell align=left, align=left, draw=white!15!black}
]
\addplot [color=mycolor1, line width=1.0pt]
  table[row sep=crcr]{%
0	4.204241380743\\
5	4.201053174599\\
10	4.19146910671\\
15	4.175579051348\\
20	4.153690348055\\
25	4.126281139297\\
30	4.093984258551\\
35	4.057567618984\\
40	4.017911140874\\
45	3.975979162872\\
50	3.932789924114\\
55	3.889384062982\\
60	3.846794556959\\
65	3.806019893215\\
70	3.768001330751\\
75	3.733604110481\\
80	3.703601889976\\
85	3.678663515866\\
90	3.65934139164\\
95	3.646061023701\\
100	3.639111749459\\
105	3.635527014627\\
110	3.632177225526\\
115	3.628992014457\\
120	3.62594209071\\
125	3.623007985822\\
130	3.620176266289\\
135	3.617436843819\\
140	3.614781917645\\
145	3.612205205795\\
150	3.609701526117\\
155	3.607266487217\\
160	3.604896291489\\
165	3.602587588597\\
170	3.60033737451\\
175	3.598142915993\\
180	3.596001696252\\
185	3.59391137388\\
190	3.591869752341\\
195	3.589874756577\\
200	3.587924415186\\
205	3.586016846555\\
210	3.584150248199\\
215	3.582322888434\\
220	3.580533100065\\
225	3.578779275535\\
230	3.5770598634\\
235	3.575373365773\\
240	3.573718336648\\
245	3.572093380775\\
250	3.570497153119\\
255	3.568928358476\\
260	3.567385751339\\
265	3.56586813562\\
270	3.564374364288\\
275	3.562903338635\\
280	3.561454007339\\
285	3.56002536507\\
290	3.558616450974\\
295	3.557226346885\\
300	3.555854175597\\
305	3.554499099073\\
310	3.553160316922\\
315	3.551837064958\\
320	3.550528614043\\
325	3.549234269032\\
330	3.547953367967\\
335	3.54668528128\\
340	3.545429411141\\
345	3.544185190717\\
350	3.542952083525\\
355	3.541729582614\\
360	3.540517209765\\
365	3.539314514527\\
370	3.538121073253\\
375	3.536936487974\\
380	3.535760385277\\
385	3.534592415071\\
390	3.533432249388\\
395	3.532279581121\\
400	3.531134122849\\
405	3.529995605633\\
410	3.528863777954\\
415	3.52773840466\\
420	3.526619266065\\
425	3.525506157056\\
430	3.524398886302\\
435	3.52329727545\\
440	3.522201158339\\
445	3.521110380126\\
450	3.520024796407\\
455	3.51894427222\\
460	3.517868681052\\
465	3.516797903789\\
470	3.515731827738\\
475	3.514670345659\\
480	3.513613354937\\
485	3.512560756833\\
490	3.511512455891\\
495	3.510468359443\\
500	3.509428377287\\
505	3.508392421435\\
510	3.507360405996\\
515	3.506332247111\\
520	3.505307862989\\
525	3.504287173955\\
530	3.50327010257\\
535	3.502256573732\\
540	3.501246514803\\
545	3.500239855717\\
550	3.499236529096\\
555	3.498236470321\\
560	3.497239617599\\
565	3.496245911997\\
570	3.49525529747\\
575	3.49426772085\\
580	3.493283131846\\
585	3.49230148301\\
590	3.491322729728\\
595	3.490346830189\\
600	3.489373745367\\
605	3.488403438998\\
610	3.487435877561\\
615	3.486471030251\\
620	3.48550886893\\
625	3.484549368045\\
630	3.483592504525\\
635	3.482638257608\\
640	3.48168660862\\
645	3.480737540695\\
650	3.479791038435\\
655	3.478847087501\\
660	3.477905674164\\
665	3.476966784752\\
670	3.47603040512\\
675	3.475096519967\\
680	3.474165112117\\
685	3.473236161787\\
690	3.472309645463\\
695	3.471385535109\\
700	3.47046379627\\
705	3.469544386931\\
710	3.468627254607\\
715	3.467712333725\\
720	3.466799540075\\
725	3.465888763672\\
730	3.464979858926\\
735	3.464072627257\\
740	3.463166778034\\
745	3.462261876623\\
750	3.461357259637\\
755	3.460451873772\\
760	3.459543810915\\
765	3.458630278415\\
770	3.457707744685\\
775	3.456771671631\\
780	3.455819554665\\
785	3.454849600558\\
790	3.453863237111\\
795	3.452867948517\\
800	3.451873112777\\
805	3.45089030962\\
810	3.449925791371\\
815	3.448978087205\\
820	3.448043297785\\
825	3.447117931485\\
830	3.446199207728\\
835	3.445284771875\\
840	3.444372256958\\
845	3.443458545065\\
850	3.442539061117\\
855	3.441607611407\\
860	3.440656288853\\
865	3.439678409517\\
870	3.438670439656\\
875	3.437629665977\\
880	3.436557737494\\
885	3.435468954637\\
890	3.434384482009\\
895	3.433325758914\\
900	3.432300528428\\
905	3.431303680511\\
910	3.430327794159\\
915	3.429366916052\\
920	3.428416313317\\
925	3.427471946903\\
930	3.426530365857\\
935	3.425589076751\\
940	3.424647683021\\
945	3.423706967514\\
950	3.422766165809\\
955	3.421822148094\\
960	3.420869957804\\
965	3.419906319801\\
970	3.418932842532\\
975	3.417953333303\\
980	3.416973987411\\
985	3.416002209367\\
990	3.415040561778\\
995	3.414086839713\\
1000	3.413137306906\\
1005	3.412189424845\\
1010	3.411241016611\\
1015	3.410288011506\\
1020	3.409324557125\\
1025	3.40834634528\\
1030	3.407353049719\\
1035	3.406345399627\\
1040	3.405329744943\\
1045	3.404317451905\\
1050	3.403315396451\\
1055	3.402324235503\\
1060	3.401341742877\\
1065	3.40036559771\\
1070	3.399394166483\\
1075	3.398425905764\\
1080	3.397458808549\\
1085	3.396489934019\\
1090	3.395514747171\\
1095	3.394526152087\\
1100	3.393514819609\\
1105	3.392471571886\\
1110	3.391395673785\\
1115	3.390296441347\\
1120	3.389179700192\\
1125	3.388040872187\\
1130	3.386866040025\\
1135	3.38563063063\\
1140	3.384304256322\\
1145	3.382884779022\\
1150	3.381403842787\\
1155	3.379923135527\\
1160	3.378502617135\\
1165	3.377159415191\\
1170	3.37588292231\\
1175	3.374657932437\\
1180	3.373471441539\\
1185	3.372313890569\\
1190	3.371177784792\\
1195	3.370057065958\\
1200	3.368946170142\\
1205	3.367839629357\\
1210	3.366731985088\\
1215	3.36561881953\\
1220	3.364498356811\\
1225	3.363373161965\\
1230	3.362247618267\\
1235	3.361124664087\\
1240	3.360003451375\\
1245	3.35888029448\\
1250	3.35774961874\\
1255	3.356603622691\\
1260	3.355434297943\\
1265	3.35423630289\\
1270	3.353004831655\\
1275	3.351738570654\\
1280	3.350445190311\\
1285	3.349130114486\\
1290	3.347792520994\\
1295	3.346432507786\\
1300	3.345052969968\\
1305	3.343667777014\\
1310	3.342292670865\\
1315	3.34093094204\\
1320	3.339577648004\\
1325	3.338225259393\\
1330	3.336869817358\\
1335	3.335520695163\\
1340	3.334192023503\\
1345	3.33289049336\\
1350	3.331614455624\\
1355	3.330357863695\\
1360	3.32911347465\\
1365	3.327873679841\\
1370	3.326631298303\\
1375	3.325380106321\\
1380	3.32411636942\\
1385	3.322837795492\\
1390	3.321541979157\\
1395	3.320225438153\\
1400	3.318883508001\\
1405	3.317509962649\\
1410	3.31609674058\\
1415	3.314633476076\\
1420	3.313103124404\\
1425	3.311481085881\\
1430	3.30973283431\\
1435	3.307831582269\\
1440	3.305798262787\\
1445	3.3037164036\\
1450	3.301685173598\\
1455	3.29975309197\\
1460	3.297919064188\\
1465	3.296164619695\\
1470	3.294471120468\\
1475	3.292822639961\\
1480	3.291204594882\\
1485	3.289602195916\\
1490	3.288002657498\\
1495	3.286396909051\\
1500	3.284777551436\\
1505	3.283149747806\\
1510	3.281533359205\\
1515	3.279947581673\\
1520	3.278400698701\\
1525	3.276892774175\\
1530	3.275420548258\\
1535	3.273979826354\\
1540	3.272566686721\\
1545	3.271177708293\\
1550	3.269810052147\\
1555	3.268461303\\
1560	3.267129456245\\
1565	3.265812856207\\
1570	3.264510168364\\
1575	3.263220255872\\
1580	3.261942149172\\
1585	3.260674985187\\
1590	3.259418006603\\
1595	3.25817052861\\
1600	3.256931942976\\
1605	3.255701695642\\
1610	3.254479290676\\
1615	3.253264274129\\
1620	3.252056237178\\
1625	3.250854804128\\
1630	3.249659634815\\
1635	3.248470415557\\
1640	3.247286860986\\
1645	3.246108707131\\
1650	3.244935712808\\
1655	3.243767654292\\
1660	3.242604326362\\
1665	3.241445538153\\
1670	3.240291113948\\
1675	3.239140889915\\
1680	3.237994714702\\
1685	3.23685244685\\
1690	3.23571395524\\
1695	3.234579117027\\
1700	3.233447817973\\
1705	3.232319950779\\
1710	3.231195415339\\
1715	3.230074117379\\
1720	3.228955968645\\
1725	3.227840885804\\
1730	3.226728790565\\
1735	3.225619608786\\
1740	3.224513270556\\
1745	3.223409709455\\
1750	3.222308862618\\
1755	3.221210670114\\
1760	3.220115074981\\
1765	3.219022022723\\
1770	3.217931461327\\
1775	3.216843340835\\
1780	3.215757613345\\
1785	3.214674232668\\
1790	3.213593154316\\
1795	3.212514335208\\
1800	3.211437733657\\
1805	3.210363309122\\
1810	3.20929102219\\
1815	3.20822083437\\
1820	3.207152708074\\
1825	3.206086606436\\
1830	3.205022493301\\
1835	3.203960333067\\
1840	3.202900090672\\
1845	3.201841731465\\
1850	3.200785221192\\
1855	3.199730525886\\
1860	3.198677611849\\
1865	3.197626445568\\
1870	3.196576993697\\
1875	3.195529222984\\
1880	3.194483100257\\
1885	3.193438592366\\
1890	3.192395666164\\
1895	3.191354288473\\
1900	3.190314426061\\
1905	3.189276045612\\
1910	3.188239113716\\
1915	3.187203596845\\
1920	3.186169461345\\
1925	3.185136673425\\
1930	3.184105199152\\
1935	3.183075004444\\
1940	3.182046055074\\
1945	3.181018316672\\
1950	3.179991754736\\
1955	3.178966334641\\
1960	3.177942021656\\
1965	3.17691878097\\
1970	3.175896577714\\
1975	3.174875376992\\
1980	3.173855143925\\
1985	3.172835843692\\
1990	3.171817441586\\
1995	3.170799903069\\
2000	3.169783193846\\
2005	3.168767279949\\
2010	3.167752127822\\
2015	3.166737704427\\
2020	3.165723977363\\
2025	3.164710915003\\
2030	3.163698486639\\
2035	3.162686662652\\
2040	3.161675414704\\
2045	3.16066471595\\
2050	3.159654541265\\
2055	3.158644867509\\
2060	3.157635673811\\
2065	3.156626941876\\
2070	3.155618656323\\
2075	3.154610805054\\
2080	3.153603379639\\
2085	3.152596375738\\
2090	3.151589793538\\
2095	3.15058363822\\
2100	3.149577920435\\
2105	3.148572656813\\
2110	3.147567870459\\
2115	3.146563591481\\
2120	3.145559857493\\
2125	3.144556714116\\
2130	3.14355421545\\
2135	3.142552424509\\
2140	3.141551413572\\
2145	3.140551264472\\
2150	3.13955206873\\
2155	3.138553927538\\
2160	3.137556951509\\
2165	3.136561260148\\
2170	3.135566980964\\
2175	3.134574248143\\
2180	3.133583200699\\
2185	3.132593979993\\
2190	3.131606726549\\
2195	3.13062157606\\
2200	3.129638654528\\
2205	3.128658072506\\
2210	3.127679918403\\
2215	3.126704250917\\
2220	3.125731090648\\
2225	3.124760410995\\
2230	3.123792128453\\
2235	3.122826092374\\
2240	3.121862074197\\
2245	3.120899756034\\
2250	3.119938718271\\
2255	3.118978425715\\
2260	3.118018211445\\
2265	3.117057257419\\
2270	3.116094570503\\
2275	3.1151289523\\
2280	3.114158960865\\
2285	3.113182861414\\
2290	3.112198561677\\
2295	3.111203524842\\
2300	3.110194642613\\
2305	3.109168016791\\
2310	3.1081184774\\
2315	3.107037975719\\
2320	3.105908918336\\
2325	3.104685606735\\
2330	3.103307026628\\
2335	3.101765387492\\
2340	3.1000763737\\
2345	3.098229249514\\
2350	3.096186911846\\
2355	3.093905956368\\
2360	3.091377364923\\
2365	3.088614736906\\
2370	3.08563100629\\
2375	3.082468601845\\
2380	3.079148763925\\
2385	3.075712465359\\
2390	3.072283652753\\
2395	3.069008232144\\
2400	3.065846091624\\
2405	3.0627908364\\
2410	3.059881239172\\
2415	3.057175952544\\
2420	3.054641449954\\
2425	3.052259974509\\
2430	3.050002405442\\
2435	3.047853697536\\
2440	3.045793861637\\
2445	3.043812356981\\
2450	3.041895262409\\
2455	3.040035353521\\
2460	3.038222544709\\
2465	3.036451709735\\
2470	3.034715160068\\
2475	3.033009083059\\
2480	3.031327348655\\
2485	3.029666997884\\
2490	3.028022978317\\
2495	3.026392965049\\
2500	3.024772796409\\
2505	3.023160766765\\
2510	3.021553560365\\
2515	3.019950157787\\
2520	3.018348152029\\
2525	3.016747268889\\
2530	3.015145869903\\
2535	3.013544099907\\
2540	3.011940588129\\
2545	3.010335310088\\
2550	3.008726331067\\
2555	3.007112612611\\
2560	3.005491118882\\
2565	3.00385924765\\
2570	3.0022119992\\
2575	3.000544360675\\
2580	2.998848241231\\
2585	2.997113269001\\
2590	2.995321250642\\
2595	2.993452998884\\
2600	2.991502698957\\
2605	2.989459854735\\
2610	2.987329366884\\
2615	2.985160132625\\
2620	2.982996844368\\
2625	2.980876696079\\
2630	2.978784614557\\
2635	2.976711107667\\
2640	2.974639945403\\
2645	2.972557694746\\
2650	2.970441585833\\
2655	2.968257333074\\
2660	2.965931223347\\
2665	2.963447626542\\
2670	2.960926703438\\
2675	2.958415953433\\
2680	2.955859631666\\
2685	2.953218323709\\
2690	2.950429522687\\
2695	2.947434143118\\
2700	2.944260608757\\
2705	2.940983749208\\
2710	2.937588796431\\
2715	2.934033806022\\
2720	2.930318245893\\
2725	2.926396734501\\
2730	2.922188222657\\
2735	2.917762355895\\
2740	2.91316461623\\
2745	2.908218413694\\
2750	2.902707150466\\
2755	2.896760607662\\
2760	2.890448434621\\
2765	2.883735356215\\
2770	2.876646699366\\
2775	2.869161419038\\
2780	2.861208538846\\
2785	2.853097697187\\
2790	2.84493036977\\
2795	2.836466917109\\
2800	2.827880329325\\
2805	2.819319752396\\
2810	2.810748496217\\
2815	2.802245230651\\
2820	2.793689760546\\
};
\addlegendentry{proposed model}

\addplot [color=mycolor2, line width=1.0pt]
  table[row sep=crcr]{%
0	4.204241380743\\
5	4.200834172749\\
10	4.190574796192\\
15	4.173538286848\\
20	4.150045824385\\
25	4.12060308783\\
30	4.085882661224\\
35	4.046702541764\\
40	4.004002011125\\
45	3.958813132288\\
50	3.912229387856\\
55	3.86537300128\\
60	3.819363055222\\
65	3.775286035428\\
70	3.734169793408\\
75	3.69696113702\\
80	3.664506769172\\
85	3.63753709876\\
90	3.616652509247\\
95	3.602311843551\\
100	3.594823075404\\
105	3.591001486314\\
110	3.587472113943\\
115	3.584138481207\\
120	3.580962621696\\
125	3.577918476721\\
130	3.574988136977\\
135	3.572158096797\\
140	3.569417972983\\
145	3.56675946314\\
150	3.564175811025\\
155	3.561661388663\\
160	3.559211436782\\
165	3.556821863449\\
170	3.554489105443\\
175	3.55221002032\\
180	3.549981807459\\
185	3.547801945902\\
190	3.545668146926\\
195	3.543578316073\\
200	3.541530523237\\
205	3.539522978262\\
210	3.537554011272\\
215	3.535622056397\\
220	3.53372563847\\
225	3.531863361926\\
230	3.530033901696\\
235	3.528235995568\\
240	3.526468437955\\
245	3.524730074644\\
250	3.523019798545\\
255	3.5213365461\\
260	3.51967929441\\
265	3.518047058764\\
270	3.516438890703\\
275	3.514853876297\\
280	3.5132911348\\
285	3.511749817405\\
290	3.510229106249\\
295	3.508728213428\\
300	3.50724638019\\
305	3.505782876057\\
310	3.50433699809\\
315	3.502908070042\\
320	3.501495441622\\
325	3.500098487644\\
330	3.498716607294\\
335	3.497349223275\\
340	3.495995781085\\
345	3.494655748205\\
350	3.493328613408\\
355	3.492013886011\\
360	3.490711095251\\
365	3.4894197896\\
370	3.488139536204\\
375	3.486869920261\\
380	3.485610544504\\
385	3.484361028618\\
390	3.483121008753\\
395	3.481890136942\\
400	3.480668080619\\
405	3.479454522043\\
410	3.478249157792\\
415	3.477051698181\\
420	3.475861866745\\
425	3.474679399646\\
430	3.473504045155\\
435	3.472335563074\\
440	3.471173724228\\
445	3.470018309912\\
450	3.468869111418\\
455	3.467725929534\\
460	3.46658857412\\
465	3.465456863659\\
470	3.464330624896\\
475	3.463209692451\\
480	3.462093908508\\
485	3.460983122475\\
490	3.459877190713\\
495	3.458775976224\\
500	3.457679348389\\
505	3.45658718266\\
510	3.455499360284\\
515	3.454415767986\\
520	3.453336297678\\
525	3.452260846124\\
530	3.451189314639\\
535	3.45012160876\\
540	3.449057637951\\
545	3.447997315299\\
550	3.446940557261\\
555	3.445887283393\\
560	3.444837416147\\
565	3.443790880663\\
570	3.442747604608\\
575	3.441707518033\\
580	3.440670553265\\
585	3.439636644802\\
590	3.438605729254\\
595	3.437577745271\\
600	3.436552633515\\
605	3.435530336618\\
610	3.434510799163\\
615	3.43349396766\\
620	3.432479790541\\
625	3.431468218139\\
630	3.430459202688\\
635	3.429452698309\\
640	3.428448661011\\
645	3.427447048678\\
650	3.426447821076\\
655	3.425450939843\\
660	3.424456368504\\
665	3.423464072459\\
670	3.422474019001\\
675	3.421486177316\\
680	3.420500518492\\
685	3.419517015519\\
690	3.418535643293\\
695	3.417556378603\\
700	3.41657920012\\
705	3.415604088364\\
710	3.414631025666\\
715	3.413659996115\\
720	3.41269098549\\
725	3.411723981176\\
730	3.410758972073\\
735	3.409795948492\\
740	3.408834902049\\
745	3.407875825543\\
750	3.406918712847\\
755	3.405963558788\\
760	3.405010359039\\
765	3.40405911001\\
770	3.403109808745\\
775	3.40216245283\\
780	3.401217040307\\
785	3.400273569586\\
790	3.399332039375\\
795	3.398392448599\\
800	3.397454796328\\
805	3.396519081699\\
810	3.395585303842\\
815	3.394653461784\\
820	3.393723554358\\
825	3.392795580091\\
830	3.391869537081\\
835	3.390945422842\\
840	3.390023234156\\
845	3.389102966857\\
850	3.388184615601\\
855	3.387268173646\\
860	3.386353632457\\
865	3.385440981387\\
870	3.384530207256\\
875	3.383621293566\\
880	3.382714220142\\
885	3.381808961918\\
890	3.380905487742\\
895	3.380003759155\\
900	3.379103726422\\
905	3.378205327378\\
910	3.377308480506\\
915	3.376413078459\\
920	3.375518972838\\
925	3.37462594884\\
930	3.373733709886\\
935	3.372841816354\\
940	3.371949598941\\
945	3.371055964136\\
950	3.370159401208\\
955	3.369257894807\\
960	3.368348668685\\
965	3.367427829616\\
970	3.366492524474\\
975	3.365543581018\\
980	3.36458581248\\
985	3.363624176733\\
990	3.362664106451\\
995	3.361712636175\\
1000	3.360773333646\\
1005	3.359845525985\\
1010	3.358926800958\\
1015	3.358014509931\\
1020	3.357106101809\\
1025	3.356198971822\\
1030	3.355290241119\\
1035	3.354376384423\\
1040	3.353453364937\\
1045	3.352516300099\\
1050	3.351558879992\\
1055	3.350575291332\\
1060	3.34956470269\\
1065	3.348533383173\\
1070	3.347491399524\\
1075	3.346452344146\\
1080	3.345429857929\\
1085	3.344429721067\\
1090	3.343450094964\\
1095	3.342486487113\\
1100	3.341534414449\\
1105	3.340589999135\\
1110	3.339649838445\\
1115	3.338710744479\\
1120	3.337769121581\\
1125	3.336820635434\\
1130	3.335861513154\\
1135	3.334889141906\\
1140	3.333901961378\\
1145	3.332899489005\\
1150	3.331881846326\\
1155	3.33085223337\\
1160	3.329817762362\\
1165	3.328784898605\\
1170	3.327757391604\\
1175	3.326736945819\\
1180	3.325726405527\\
1185	3.324728201321\\
1190	3.323741925978\\
1195	3.322764778722\\
1200	3.321793354504\\
1205	3.320825353969\\
1210	3.319860185814\\
1215	3.318898529054\\
1220	3.317941479066\\
1225	3.316989846198\\
1230	3.316043846468\\
1235	3.315103142375\\
1240	3.314167066318\\
1245	3.313234799166\\
1250	3.312305382029\\
1255	3.311377729367\\
1260	3.310450588839\\
1265	3.30952249213\\
1270	3.308591624778\\
1275	3.307655644166\\
1280	3.306711514115\\
1285	3.305755228444\\
1290	3.304781168101\\
1295	3.303781889648\\
1300	3.302747491303\\
1305	3.30166437024\\
1310	3.300515431019\\
1315	3.299290344331\\
1320	3.297992348441\\
1325	3.29663416683\\
1330	3.295237220678\\
1335	3.293842832165\\
1340	3.292492042439\\
1345	3.291200735519\\
1350	3.289964782656\\
1355	3.288773374693\\
1360	3.287616137987\\
1365	3.286484672667\\
1370	3.285372280684\\
1375	3.284273402719\\
1380	3.283183244181\\
1385	3.282097253354\\
1390	3.281010797948\\
1395	3.27992003218\\
1400	3.278821546075\\
1405	3.277711430866\\
1410	3.276584978931\\
1415	3.275437765275\\
1420	3.274266334109\\
1425	3.273067529514\\
1430	3.271842769753\\
1435	3.270597373716\\
1440	3.269335959205\\
1445	3.268059455144\\
1450	3.266764870679\\
1455	3.265453199559\\
1460	3.264132032743\\
1465	3.262808214022\\
1470	3.261484322875\\
1475	3.260159995637\\
1480	3.258834727558\\
1485	3.257513416289\\
1490	3.256210340546\\
1495	3.254938596253\\
1500	3.253702687115\\
1505	3.252500283747\\
1510	3.251326535722\\
1515	3.250176648546\\
1520	3.249046557354\\
1525	3.247932960315\\
1530	3.246833176081\\
1535	3.245745004459\\
1540	3.244666593547\\
1545	3.243596344734\\
1550	3.242532836262\\
1555	3.241474754731\\
1560	3.240420835267\\
1565	3.239369806156\\
1570	3.238320261424\\
1575	3.237270543881\\
1580	3.236218651331\\
1585	3.235162096727\\
1590	3.234097866023\\
1595	3.23302218032\\
1600	3.231930252581\\
1605	3.230816804895\\
1610	3.22967677284\\
1615	3.228505363206\\
1620	3.227297440028\\
1625	3.226046903389\\
1630	3.224746791793\\
1635	3.223389337164\\
1640	3.221965295111\\
1645	3.220464254925\\
1650	3.218875990578\\
1655	3.217196135943\\
1660	3.215434569253\\
1665	3.213620852989\\
1670	3.211797156308\\
1675	3.209994525755\\
1680	3.208225511687\\
1685	3.206490191492\\
1690	3.20478161712\\
1695	3.203098100946\\
1700	3.201447491617\\
1705	3.199839082433\\
1710	3.198277217492\\
1715	3.196761446676\\
1720	3.195288382151\\
1725	3.193853538994\\
1730	3.192452439419\\
1735	3.191081033454\\
1740	3.189735846873\\
1745	3.188413915807\\
1750	3.187112700052\\
1755	3.185830034678\\
1760	3.184564136788\\
1765	3.18331353479\\
1770	3.182076990992\\
1775	3.180853434467\\
1780	3.179641924956\\
1785	3.17844162609\\
1790	3.177251790794\\
1795	3.176071747526\\
1800	3.174900889643\\
1805	3.173738665806\\
1810	3.172584573717\\
1815	3.171438154334\\
1820	3.170298987225\\
1825	3.169166686334\\
1830	3.168040896762\\
1835	3.166921291337\\
1840	3.165807568055\\
1845	3.164699447573\\
1850	3.163596671196\\
1855	3.162498998918\\
1860	3.161406207777\\
1865	3.160318090321\\
1870	3.159234453285\\
1875	3.158155116362\\
1880	3.157079911117\\
1885	3.156008679998\\
1890	3.154941275436\\
1895	3.153877559044\\
1900	3.152817400865\\
1905	3.15176067872\\
1910	3.150707277576\\
1915	3.149657089008\\
1920	3.148610010674\\
1925	3.147565945861\\
1930	3.146524803044\\
1935	3.145486495516\\
1940	3.144450941004\\
1945	3.143418061363\\
1950	3.142387782252\\
1955	3.14136003287\\
1960	3.14033474569\\
1965	3.139311856228\\
1970	3.138291302813\\
1975	3.137273026401\\
1980	3.136256970371\\
1985	3.135243080366\\
1990	3.134231304121\\
1995	3.133221591325\\
2000	3.13221389347\\
2005	3.131208163741\\
2010	3.130204356878\\
2015	3.129202429081\\
2020	3.128202337896\\
2025	3.12720404213\\
2030	3.126207501748\\
2035	3.125212677809\\
2040	3.124219532367\\
2045	3.123228028421\\
2050	3.122238129825\\
2055	3.12124980125\\
2060	3.1202630081\\
2065	3.119277716479\\
2070	3.118293893122\\
2075	3.117311505362\\
2080	3.116330521071\\
2085	3.115350908633\\
2090	3.114372636887\\
2095	3.113395675107\\
2100	3.112419992955\\
2105	3.111445560457\\
2110	3.110472347963\\
2115	3.10950032613\\
2120	3.108529465882\\
2125	3.107559738396\\
2130	3.10659111507\\
2135	3.105623567507\\
2140	3.104657067487\\
2145	3.103691586954\\
2150	3.102727097991\\
2155	3.101763572808\\
2160	3.10080098372\\
2165	3.09983930314\\
2170	3.098878503553\\
2175	3.097918557516\\
2180	3.096959437631\\
2185	3.096001116548\\
2190	3.095043566946\\
2195	3.094086761527\\
2200	3.093130673006\\
2205	3.092175274097\\
2210	3.091220537525\\
2215	3.090266436003\\
2220	3.089312942236\\
2225	3.088360028917\\
2230	3.087407668724\\
2235	3.086455834326\\
2240	3.085504498377\\
2245	3.084553633525\\
2250	3.083603212411\\
2255	3.082653207687\\
2260	3.081703592012\\
2265	3.080754338077\\
2270	3.079805418607\\
2275	3.078856806389\\
2280	3.077908474285\\
2285	3.076960395259\\
2290	3.0760125424\\
2295	3.075064888958\\
2300	3.07411740837\\
2305	3.073170074303\\
2310	3.072222860698\\
2315	3.071275741803\\
2320	3.070328692233\\
2325	3.069381687018\\
2330	3.068434701659\\
2335	3.067487712185\\
2340	3.066540695213\\
2345	3.065593628018\\
2350	3.06464648859\\
2355	3.063699255705\\
2360	3.062751908994\\
2365	3.061804429009\\
2370	3.060856797293\\
2375	3.059908996448\\
2380	3.058961010196\\
2385	3.058012823456\\
2390	3.057064422396\\
2395	3.056115794503\\
2400	3.055166928633\\
2405	3.054217815073\\
2410	3.053268445582\\
2415	3.052318813437\\
2420	3.051368913469\\
2425	3.050418742082\\
2430	3.049468297269\\
2435	3.048517578604\\
2440	3.047566587216\\
2445	3.046615325736\\
2450	3.045663798214\\
2455	3.044712009996\\
2460	3.043759967556\\
2465	3.042807678267\\
2470	3.041855150108\\
2475	3.040902391292\\
2480	3.039949409798\\
2485	3.038996212806\\
2490	3.038042806005\\
2495	3.037089192776\\
2500	3.036135373223\\
2505	3.035181343051\\
2510	3.034227092275\\
2515	3.033272603754\\
2520	3.032317851534\\
2525	3.031362799017\\
2530	3.030407396938\\
2535	3.02945158116\\
2540	3.028495270287\\
2545	3.027538363095\\
2550	3.026580735776\\
2555	3.025622239\\
2560	3.024662694747\\
2565	3.023701892897\\
2570	3.022739587525\\
2575	3.021775492807\\
2580	3.020809278477\\
2585	3.019840564673\\
2590	3.01886891607\\
2595	3.017893835085\\
2600	3.016914753984\\
2605	3.015931025638\\
2610	3.014941912638\\
2615	3.013946574495\\
2620	3.012944052408\\
2625	3.011933251353\\
2630	3.010912918502\\
2635	3.009881617754\\
2640	3.008837698257\\
2645	3.007779256859\\
2650	3.006704088682\\
2655	3.005609620824\\
2660	3.004492811524\\
2665	3.003349974659\\
2670	3.002176401013\\
2675	3.000965409573\\
2680	2.999705802808\\
2685	2.998374846261\\
2690	2.996925551065\\
2695	2.995275052894\\
2700	2.993335246898\\
2705	2.9910912341\\
2710	2.98856292556\\
2715	2.985767940441\\
2720	2.982731978904\\
2725	2.979488804411\\
2730	2.976088922062\\
2735	2.972671918637\\
2740	2.969362177389\\
2745	2.966231277486\\
2750	2.963313012216\\
2755	2.960599515825\\
2760	2.958069972083\\
2765	2.955698992927\\
2770	2.953462650789\\
2775	2.951340561689\\
2780	2.949315382255\\
2785	2.94737304484\\
2790	2.94550174664\\
2795	2.943691767339\\
2800	2.9419347846\\
2805	2.940223775788\\
2810	2.938552600806\\
2815	2.936915965465\\
2820	2.935309158508\\
2825	2.93372803668\\
2830	2.932168856497\\
2835	2.930628260939\\
2840	2.92910316456\\
2845	2.927590742036\\
2850	2.926088348298\\
2855	2.924593503454\\
2860	2.923103837694\\
2865	2.92161708052\\
2870	2.920131021944\\
2875	2.918643502581\\
2880	2.917152388035\\
2885	2.915655561967\\
2890	2.914150910792\\
2895	2.912636317696\\
2900	2.911109651989\\
2905	2.90956877384\\
2910	2.908011553179\\
2915	2.906435881316\\
2920	2.904839671562\\
2925	2.90322086612\\
2930	2.901577480707\\
2935	2.899907602045\\
2940	2.898209308816\\
2945	2.896480576411\\
2950	2.894719032724\\
2955	2.892921584065\\
2960	2.891084145597\\
2965	2.889201296338\\
2970	2.887265992516\\
2975	2.885269508703\\
2980	2.883203589357\\
2985	2.881067441011\\
2990	2.878876769465\\
2995	2.876657157662\\
3000	2.874426148431\\
3005	2.872198954007\\
3010	2.8699948903\\
3015	2.867821439778\\
3020	2.86566947654\\
3025	2.863527634351\\
3030	2.861386857606\\
3035	2.859238278421\\
3040	2.857071879004\\
3045	2.854874364179\\
3050	2.852623623807\\
3055	2.850278020401\\
3060	2.847785732353\\
3065	2.845131203637\\
3070	2.842340679082\\
3075	2.839474012717\\
3080	2.836566140024\\
3085	2.833644866964\\
3090	2.830762001573\\
3095	2.827944490352\\
3100	2.825191027973\\
3105	2.822482599641\\
3110	2.819800268159\\
3115	2.817126964705\\
3120	2.814448152452\\
3125	2.811749892232\\
3130	2.809018741917\\
3135	2.806240526082\\
3140	2.80339995796\\
3145	2.800479210463\\
3150	2.797455474656\\
};
\addlegendentry{0D interface model}

\end{axis}
\end{tikzpicture}%

%% file: chapters/chapter_5.tex
We presented a novel modeling approach to incorporate the transport of mass and charge in the tangential direction of coating layers in ASSBs embedded into an electrochemo-mechanical model for three-dimensionally resolved microstructures. By spatially resolving the thin coating layer only in two dimensions of space and assuming the normal component of the electric field being constant along the normal direction, an extremely fine resolution of the mesh can be avoided, which so far was the limiting factor to consider transport in the tangential direction of coating layers in realistic microstructures of ASSBs. The outlined model is numerically discretized in space using the finite element method and the resulting system of algebraic equations is solved in a monolithic fashion considering both the different physical fields and the different geometric entities, i.e. surfaces and volumes. We justified the assumptions we made by analyzing the result of a fully resolved simulation, validated the implementation by showing the conservation of mass and quadratic convergence of the error, and discussed the implications of our assumptions. \\
Finally, numerical examples are presented to show the applicability of the novel modeling approach to realistic microstructures and to outline that established models which capture the coating layer by an additional zero-dimensional resistance can lead to significant errors. By a systematic variation of the most influential parameters of a coating layer, a deeper understanding of the physics inside a coating layer and its influence on the cell performance was gained. We showed that different coating strategies are advantageous depending on the operating scenarios (i.e. the boundary conditions), and incorporating conduction along the coating layer into simulation models is essential in finding the optimal parameters of the coating layer.\\
In future research, the novel model can be applied to geometrically similar domains like grain boundaries in the solid electrolyte where a priori knowledge on the development of quantities along the thin direction is given or can be derived.

%% file: chapters/funding.tex
We gratefully acknowledge support by the Bavarian Ministry of Economic Affairs, Regional Development and Energy [project ``Industrialisierbarkeit von Festk\"orperelektrolytzellen''] and the German Federal Ministry of Education and Research [FestBatt~2 (03XP0435B)].

%% file: chapters/appendix.tex
\section*{List of symbols}
\setcounter{table}{4}
\renewcommand{\thetable}{\arabic{table}}
\begin{table}[H]
    \begin{tabular}{c | l}
        \hline
        \multicolumn{2}{l}{\textbf{Geometric quantities}} \\
        \hline
        $l_i$ & length of domain $i$ \\
        $\vec{n}$ & unit normal vector \\
        $t_\text{coat}$ & thickness of coating layer \\
        $t_i,n$ & coordinates inside of coating layer \\
        $\vec{X}$ & material coordinate \\
        $\vec{x}$ & spatial coordinate \\
        $\Gamma_{i-j}$ & intersection of domains $i$ and $j$ \\
        $\Omega_i$ & domain $i$ \\   
        \hline
        \multicolumn{2}{l}{\textbf{Constants}} \\
        \hline
        $F$ & Faraday constant \\
        $R$ & universal gas constant \\
        \hline
        \multicolumn{2}{l}{\textbf{Model parameters}} \\
        \hline
        $c_*$ & concentration with identifier $*$ \\
        $D$ & diffusion coefficient \\
        $\mat{F}_*$ & deformation gradient with identifier $*$\\
        $\vec{i}$ & current density of charge \\
        $i_0$ & exchange current density \\
        $\vec{j}$ & flux density of mass \\
        $\mat{K}_{\Psi_1, \Psi_2}^i$ & derivative of residual of quantity~$\Psi_1$ w.r.t. quantity~$\Psi_2$ in domain~$i$ \\
        $\vec{R}_\Psi^i$ & residual of quantity $\Psi$ in domain $i$ \\
        $r_\text{i}$ & interface resistance \\
        $s_\Psi$ & source term of quantity $\Psi$ \\
        $T$ & temperature \\
        $t$ & time \\
        $t_+$ & transference number \\
        $W_*$ & energy with identifier $*$ \\
        $z$ & charge number \\
        $\epsilon$ & electronic conductivity \\ 
        $\lambda_*$ & Lagrange multiplier with identifier $*$ \\
        $\Phi$ & electric potential \\
        $\Psi, \vec{\Psi}$ & exemplary quantity \\
        $\sigma$ & electronic conductivity \\
        $\rho$ & mass density \\
        $\kappa$ & ionic conductivity \\ 
        \hline
    \end{tabular}
    \caption{List of symbols.}
    \label{table:list_of_symbols}
\end{table}
\setcounter{table}{0}
\renewcommand{\thetable}{\Alph{section}.\arabic{table}}
\section{Rough estimation of computational efficiency}
\label{sec:rough_estimate}
The geometry shown in \cref{fig:realisitc_geometry} consists of 564,711 bulk nodes and 71,785 nodes at the interface between the solid electrolyte and the cathode where the coating layer is added and an average edge size of the elements of $l_\text{nov} = 0.4 \ \mu \text{m}$. With the novel approach, the number of unknowns is~$n_\text{nov} = 564,711 \cdot 5 + 71,785 \cdot 2 = 2,967,125$ (bulk: displacements in three directions, concentration, and electric potential; coating layer: concentration and electric potential as unknowns per node).\\
Assuming 3 elements in the direction of the thickness and an element aspect ratio of the elements of 5, a resolved model would require a spatial discretization of the coating layer with a desired edge length in the tangential direction of the elements of $l_\text{res} = \frac{t_\text{coat}}{3} \cdot 5 = 0.0167 \ \mu \text{m}$, with $t_\text{coat} = 10 \ \text{nm}$. Thus, a refinement of the nodes at the interface of $\left(\frac{0.4 \ \mu \text{m}}{0.0167 \ \mu \text{m}}\right)^2 \cdot 3 \approx 1875$ would be required. This means that $n_\text{nov} = 564,711 \cdot 5 + 71,785 \cdot 2 \cdot 1875 = 272,017,305$ unknowns would be necessary to spatially resolve the coating layer. Thus, the proposed model reduces the number of unknowns by two orders of magnitude. The reduction of the computational time is expected to be even larger as many solvers do not scale linearly with the numbers of unknowns. Additionally, a fine mesh would also be required in the adjacent bulk regions if coupling algorithms for non-matching discretizations are not available. This would probably add at least another order of magnitude of unknowns to the resolved case.
\section{Governing equations for the bulk domain in a continuum formulation}
\label{sec:bulk_model}
The governing equations for the conservation of momentum, charge, and mass, as well as the coupling constraints between the fields of solid mechanics and electrochemistry are dervied in the following (see~\cite{Schmidt2022} for details).
\subsection{Solid Mechanics}
The equations of solid mechanics are formulated geometrically nonlinear by introducing a reference configuration (where quantities are denoted by capital letters) and a current configuration (where quantities are denoted by small letters). With~$\vec{X}$ and~$\vec{x}$ being the coordinates in reference and current configuration, respectively, and by defining $\nabla_{\vec{X}} \cdot$ as the divergence in the reference configuration, $\mat{F} = \parder{\vec{x}}{\vec{X}}$ as the deformation gradient, $\mat{S}$ as the second Piola-Kirchhoff stress tensor, $\rho_0$ as the mass density in the reference configuration, $\vec{u} = \vec{x} - \vec{X}$ as the displacement, and $\vec{b}_0$ as external volumetric loads, the balance of linear momentum is given as 
\begin{equation}
    \nabla_\mathbf{X} \cdot (\mat{F} \mat{S}) + \vec{b}_0 = \rho_0 \ddot{\vec{u}} \quad \text{in} \ \Omega_0.
\end{equation}
The deformation can be split into different contributions by making use of the chain rule for the deformation gradient to model a purely elastic deformation~$\mat{F}_\text{el}$ and another deformation~$\mat{F}_\text{growth}$ arising from stress-free volumetric growth or shrinkage due to (de-)lithiation of the electrodes
\begin{equation}
    \mat{F} = \parder{\vec{x}}{\vec{X}} = \parder{\vec{x}}{\vec{X}_1}  \parder{\vec{X}_1}{\vec{X}} = \mat{F}_\text{el} \mat{F}_{\text{growth}}.
\end{equation}
The mechanical stresses are induced by the elastic part of the deformation and follow a hyperelastic material law. The stresses are transformed to the reference configuration by a pull-back operation
\begin{equation}
    \mat{S} = 2 \ \text{det}(\mat{F}_\text{growth}) \mat{F}_\text{growth}^{-1} \parder{\Psi_\text{el}}{\mat{C}_\text{el}} \mat{F}_\text{growth}^{-\text{T}},
\end{equation}
where $\Psi_\text{el}$ is any strain energy function of a hyperelastic material law (the \textit{Neo-Hooke} material law \cite{Holzapfel2000} is exemplarily used in this work) and $\mat{C}_\text{el} = \mat{F}_\text{el}^\text{T} \mat{F}_\text{el}$. The conservation of angular momentum is implicitly fulfilled due to the symmetry of the stress tensor $\mat{S}$.
\subsection{Electrochemistry}
Both conservation of mass and conservation of charge are derived from the general form of conservation for a volume-specific quantity~$\Psi = \{c, \rho \}$, with the concentration $c$ of the transported species (lithium in the electrodes and lithium-ions in the electrolyte) and the charge density~$\rho$  (electrons in the electrodes and current collectors and lithium-ions in the electrolyte), the flux density of the respective quantity~$\vec{j}_\Psi$, and the volumetric source of this quantity~$s_\Psi$
\begin{equation}
    \parder{\Psi}{t} + \nabla \cdot \vec{j}_\Psi = s_\Psi.
\end{equation}
Assuming electro-neutrality within the entire domain, and subsequently, no accumulation of free charge ($\parder{\rho}{t}=0$) and no sources of charge ($s_\rho=0$), the conservation of charge can be expressed as $\nabla \cdot \vec{i} = 0$, with $\vec{i} = \vec{j}_\rho$ the electric current density. Inside the electrodes and the current collectors, electrons are assumed to be the only mobile charge carriers, such that the current density follows Ohm's law $\vec{i} = - \sigma \nabla \Phi$, with the electronic conductivity~$\sigma$ resulting in the Laplace equation for the electric potential $\Phi$
\begin{equation}    
    \nabla \cdot (- \sigma \nabla \Phi) = 0 \quad \text{in} \ \Omega_\text{ed} \cup \Omega_\text{cc}.
\end{equation}
The flux density of ions inside the solid electrolyte is described by the Nernst-Planck equation, i.e.~convection, migration, and diffusion. As common for many solid electrolytes, only one species of ions is modeled as mobile. In combination with the condition of local electro-neutrality, the only remaining transport effect is migration resulting in $\vec{i} = - \kappa \nabla \Phi$, with the ionic conductivity~$\kappa$. Finally, this leads to the Laplace equation for the electric potential in the solid electrolyte
\begin{equation}
    \label{eq:pot_solid_electrolyte_appendix}
    \nabla \cdot (- \kappa \nabla \Phi) = 0 \quad \text{in} \ \Omega_\text{el}.
\end{equation}
No sources of mass occur ($s_c=0$) when considering the conservation of mass. In the electrodes, only electrons are charge carriers, such that all species with mass have a charge number of zero, and thus, the migration term vanishes. This results in the conservation of mass for the electrodes
\begin{equation}
    \left.\parder{c}{t}\right|_{\vec{X}} + c \ \nabla \cdot \dot{\vec{u}} - \nabla \cdot (D \nabla c) = 0 \quad \text{in} \ \Omega_\text{ed},
\end{equation}
where $\left.\parder{c}{t}\right|_{\vec{X}}$ denotes the material time derivative, $D$ the diffusion coefficient, and~$\dot{\vec{u}}$ the deformation velocity of the domain. In the solid electrolyte, diffusion vanishes as no gradients of the concentration arise due to the electro-neutrality condition and a transference number of one. Additionally, the migration term vanishes, as it evaluates to zero (see \cref{eq:pot_solid_electrolyte_appendix}) leading to
\begin{equation}
    \left.\parder{c}{t}\right|_{\vec{X}} + c \ \nabla \cdot \dot{\vec{u}} = 0 \quad \text{in} \ \Omega_\text{el}.
\end{equation}
In this work, we model the coating layer as an ion conductor, such that all equations that are applied to the solid electrolyte domain~$\Omega_\text{el}$ are also applied to the coating domain~$\Omega_\text{coat}$. However, the presented approach can also be applied to different models for the transport of mass and charge in the coating domain, which would be relevant in e.g. co-conductive coating layers (e.g. \cite{Wang2016,Shim2016}).
\subsection{Coupling between solid mechanics and electrochemistry}
We model the coupling between solid mechanics and electrochemistry in both directions: The displacement and velocity computed from the equations of solid mechanics define the deformation of the underlying geometry of the electrochemical equations. The part of the deformation gradient originating from growth is modeled by $\mat{F}_\text{growth} = \textbf{fn}(c)$ in the electrodes~$\Omega_\text{ed}$ to account for the lithiation dependent volume change of the electrodes (see e.g.~\cite{Koerver2018} for $\textbf{fn}(c)$ for various electrode materials). In the other domains ($\Omega \ \backslash \ \Omega_\text{ed}$), no lithiation-dependent growth occurs, such that we model $\mat{F}_\text{growth} = \mat{1}$.
\section{Scalar transport equations on curved surfaces}
\label{sec:curved_nabla}
Transport equations on curved surfaces have already been discussed and applied to various applications in the literature, e.g. \cite{Rosenberg2003} for the Laplace equation, \cite{Schwartz2005, Sbalzarini2006} for pure diffusion, or \cite{Vuong2017} for diffusion-reaction. \\
The formulation of transport equations on curved surfaces requires a modification of the derivative operators, as the direction of the change of a quantity can only be along the curved surface. The gradient of a scalar~$\Phi$ w.r.t. a curved surface~$\Gamma$, denoted as~$\nabla_\Gamma$, is given by the classical gradient in the tangent plane direction~$\nabla \Phi$ corrected by changes in the direction out of the plane which depend on the local curvature of the surface, i.e.
\begin{equation}
    \nabla_\Gamma \phi = \left( \mat{1} - \vec{n} \otimes \vec{n} \right) \nabla \phi,
\end{equation}
with the unit normal vector $\vec{n}$ of a surface. The general form of the divergence of a vector~$\vec{\psi}$ is defined as
\begin{equation}
    \nabla_\Gamma \cdot \vec{\psi} = \frac{1}{\sqrt{\text{det}(\mat{g})}} \sum_i \parder{\sqrt{\text{det}(\mat{g})} \psi_i}{x_i},
\end{equation}
with the metric tensor~$\mat{g}$, which equals the identity matrix in the Euclidean space with Cartesian coordinates. Now, the transport equations on curved surfaces can be reformulated as
\begin{equation}
    \parder{\Psi}{t} + \nabla_\Gamma \cdot \vec{j}_\Psi = s,
\end{equation}
with the flux density $\vec{j}_\Psi$ of a scalar~$\Psi$.
\section{Material parameters}
\label{sec:mat_params}
The material parameters used for the simulations for the comparison with a resolved model are summarized in \cref{table:material_parameters_validation}. These parameters are artificially selected to stress differences between a resolved model and the presented model and to highlight the influence of the assumptions we made.
\input{tables/material_parameters_validation.tex} \noindent
The material parameters used for the simulations with the realistic geometry are summarized in \cref
{table:material_parameters_realistic}.
\input{tables/material_parameters_realistic.tex} \noindent
The conductivity
\begin{equation}
    \sigma(x) = 100 \frac{\text{S}}{\text{m}} \, \exp(-202.90 \, x^4 + 322.38 \, x^3 - 178.23 \, x^2 + 50.06 \, x - 13.47), \label{eq:nmc_cond}
\end{equation}
with $x = 1 - \chi$ and $\chi = \frac{c}{c_\text{max}} \chi_\text{max} \text{det}(\mat{F})$, and the diffusion coefficient of NMC622 are a function of the lithiation state~\cite{Neumann2020}
\begin{equation}
    \begin{split}
        D(\chi) &= \frac{1}{1000} \frac{\text{m}^2}{\text{s}} \exp \Big(9.3764575854 \cdot 10^5 \cdot \chi^9 - 5.4262087319 \cdot 10^6 \cdot \chi^8 \\
        &+ 1.3688556703 \cdot 10^7 \cdot \chi^7 - 1.9734363260 \cdot 10^7 \cdot \chi^6 + 1.7897244160 \cdot 10^7 \cdot \chi^5 \\
        & - 1.0576735297 \cdot 10^7 \cdot \chi^4 + 4.0688465295 \cdot 10^6 \cdot \chi^3 - 9.8167452940 \cdot 10^5 \cdot \chi^2 \\
        &+ 1.3468923578 \cdot 10^5 \cdot \chi - 8.0270847914 \cdot 10^3 \Big) \, .
    \end{split} \label{eq:nmc_diff}
\end{equation}
The open circuit potential~\cite{Kremer2019} of NMC622 is shown in~\cref{fig:OCP_NMC622}.
\begin{figure}[H]
    \centering
    \input{figures/OCP_NMC622.tikz}
    \caption{Open circuit potential of NMC622 as a function of the lithiation state based on~\cite{Kremer2019}.}
    \label{fig:OCP_NMC622}
\end{figure}
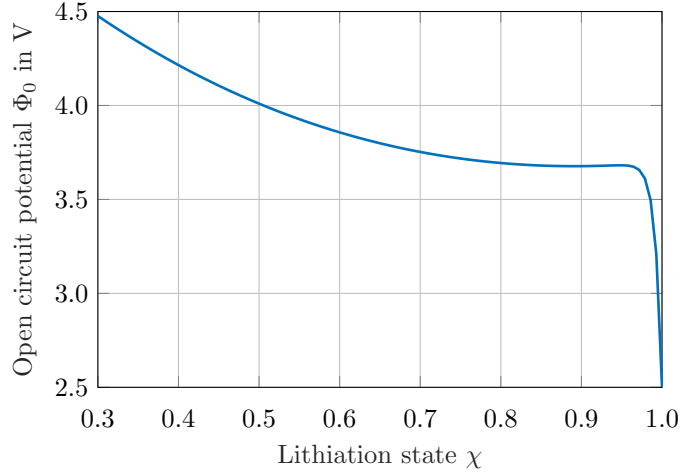 \noindent
The growth laws for the electrodes are a function of the concentration
\begin{align}
    \mat{F}_\text{growth} & = \mat{1} + \left[g \ \text{det}(\mat{F}) \frac{ \left( n_\text{ed} - n_\text{ed}^0 \right) }{V} \right] \vec{g} \otimes \vec{g} \quad \text{in} \ \Omega_\text{a}, \label{eq:growth_anode}\\
    \mat{F}_\text{growth} & = \left( \frac{ f(\chi) + 1}{f(\chi_0) + 1} \right)^{1/3} \mat{1} \quad \text{in} \ \Omega_\text{c}, \label{eq:growth_cathode}
\end{align}
with $g = \frac{M}{\rho} = 1.2998 \cdot 10^{-5} \ \frac{\text{m}^3}{\text{mol}}$ \cite{Schmidt2022}, $n_\text{ed} = \int_{\Omega_\text{ed}} c \ \text{d}v$, $\vec{g}$ the direction of growth, and $f(\chi) = \sum_{n=0}^7 \alpha_n \chi^n = 0.000444577043098 - 1.24116361022373 \chi + 9.30461909734883 \chi^2 - 29.44977325195 \chi^3 + 49.1126838772603 \chi^4 - 45.1097641074935 \chi^5 + 21.5994362668471 \chi^6 - 4.21656846170118 \chi^6$~\cite{Koerver2018}.

%% file: tables/material_parameters_validation.tex
\begin{table}[H]
    \renewcommand{\arraystretch}{1.1}
    \centering
    \caption{Material parameters for the comparison with a resolved model.}
    \begin{tabular}{c | c | c | c  }
      \hline
      \textbf{domain}  
                                                &      \textbf{quantity}   & \textbf{symbol}                        & \textbf{value} \\
      \hline
      \multirow{3}{*}{cathode $\Omega_\text{c}$}                 & diffusion coefficient    & $D$                                    & $1.83 \cdot 10^{-14} \frac{\text{m}^2}{\text{s}}$                                 \\
                                                                 & growth law               & $\mat{F}_\text{growth}$                                  & $\mat{1}$                          \\
                                                                 & open circuit potential   & $\Phi_0$                               & $0 \ \text{V}$                        \\
      \hline
      \multirow{2}{*}{coating $\Omega_\text{coat}$}              & ionic conductivity       & $\kappa$                               & $\{10^{-4}, 10^{-3}, 10^{-2}\} \cdot 5 \ \frac{\text{S}}{\text{m}}$    \\
                                                                 & transference number      & $t_+$                                    & 1                                                 \\
      \hline
      \multirow{2}{*}{solid electrolyte $\Omega_\text{el}$}      & ionic conductivity       & $\kappa$                               & $5 \cdot 10^{-4} \ \frac{\text{S}}{\text{m}}$    \\
                                                                 & transference number      & $t_+$                                    & 1                                                 \\
      \hline
      \shortstack{interface cathode - \\ coating $\Gamma_\text{coat-c}$}                 & exchange current density & $i_0$                                  & $4.98 \ \frac{\text{A}}{\text{m}^2}$    \\
      \hline
      \shortstack{interface coating - \\ solid electrolyte $\Gamma_\text{el-coat}$}      & interface resistance     & $r_\text{i}$                                    & $5.0 \cdot 10^{-3} \ \Omega \text{m}^2$               \\
      \hline
    \end{tabular}
    \label{table:material_parameters_validation}
\end{table}

%% file: tables/material_parameters_realistic.tex
\begin{table}[H]
    \renewcommand{\arraystretch}{1.1}
    \centering
    \caption{Material parameters for the simulation with the realistic geometry.}
    \begin{tabular}{c | c | c | c | c }
      \hline
      \textbf{domain}                                            &      \textbf{quantity}   & \textbf{symbol}                        & \textbf{value}                                    & \textbf{source} \\
      \hline
      \multirow{10}{*}{cathode $\Omega_\text{c}$}                & electronic conductivity  & $\sigma$                               & \cref{eq:nmc_cond}                                & \cite{Neumann2020} \\
                                                                 & diffusion coefficient    & $D$                                    & \cref{eq:nmc_diff}                                & \cite{Neumann2020} \\
                                                                 & open circuit potential   & $\Phi_0$                               & \cref{fig:OCP_NMC622}                             & \cite{Kremer2019} \\
                                                                
                                                                 & max. concentration       & $c_\text{max}$                         & $5.19 \cdot 10^4 \ \frac{\text{mol}}{\text{m}^3}$ & \cite{Neumann2020} \\
                                                                 & max. lithiation          & $\chi_\text{max}$                      & $1$                                               & \cite{Neumann2020} \\
                                                                 & lithiation range         & $[\chi_\text{0\%}, \chi_\text{100\%}]$ & $[1, 0.404]$                                      & \cite{Schmidt2022} \\
                                                                 & Young's modulus          & $E$                                    & $1.78 \cdot 10^{11} \ \text{Pa}$                  & \cite{Sun2017} \\
                                                                 & Poisson's ratio          & $\nu$                                  & $0.3$ & \cite{Xu2017}\\
                                                                 & growth law               & $\mat{F}_\text{growth}$                                  & \cref{eq:growth_cathode}                          & \cite{Schmidt2022}, based on \cite{Koerver2018} \\
      \hline
      \multirow{5}{*}{coating $\Omega_\text{coat}$}              & ionic conductivity       & $\kappa$                               & $5 \cdot 10^{-4} \ \frac{\text{S}}{\text{m}}$     & \cite{Javed2020}, based on \cite{Glass1978} \\
                                                                 & transference number      & $t_+$                                    & 1                                                 & defined, based on \cite{Wang2015} \\
                                                                 & Young's modulus          & $E$                                    & $247.6 \cdot 10^9 \ \text{Pa}$                    & \cite{Yang2013} \\
                                                                 & Poisson's ratio          & $\nu$                                  & $0.13$ & \cite{Yang2013}\\
                                                                 & ion concentration        & $c_\text{coat,0}$                            & $1.03 \cdot 10^4 \frac{\text{mol}}{\text{m}^3}$   & adapted from $\Omega_\text{el}$ \\
      \hline
      \multirow{5}{*}{solid electrolyte $\Omega_\text{el}$}     & ionic conductivity       & $\kappa$                               & $1.2 \cdot 10^{-2} \ \frac{\text{S}}{\text{m}}$   & \cite{Neumann2020} \\
                                                                 & transference number      & $t_+$                                    & 1                                                 & \cite{Neumann2020} \\
                                                                 & Young's modulus          & $E$                                    & $2.89 \cdot 10^{10} \ \text{Pa}$                  & \cite{Koerver2018} \\
                                                                 & Poisson's ratio          & $\nu$                                  & $0.27$                                            & \cite{Koerver2018} \\
                                                                 & ion concentration        & $c_\text{el,0}$                            & $1.03 \cdot 10^4 \frac{\text{mol}}{\text{m}^3}$   & \cite{Neumann2020} \\
      \hline
      \multirow{4}{*}{anode $\Omega_\text{a}$}                  & electronic conductivity  & $\sigma$                               & $10^5 \ \frac{\text{S}}{\text{m}}$                & \cite{Neumann2020} \\
                                                                 & Young's modulus          & $E$                                    & $4.9 \cdot 10^9 \ \text{Pa}$                      & \cite{Koerver2018} \\
                                                                 & Poisson's ratio          & $\nu$                                  & $0.42$                                            & \cite{Koerver2018} \\
                                                                 & growth law               & $\mat{F}_\text{growth}$                                  & \cref{eq:growth_anode}                            & \\
      \hline
      \multirow{3}{*}{\shortstack{current collector \\ anode $\Omega_\text{ac}$}}  & electronic conductivity  & $\sigma$                               & $5.81 \cdot 10^7 \ \frac{\text{S}}{\text{m}}$     & \cite{JensFreudenberger2020} \\
                                                                 & Young's modulus          & $E$                                    & $1.15 \cdot 10^{11} \ \text{Pa}$                  & \cite{Freund2004} \\
                                                                 & Poisson's ratio          & $\nu$                                  & $0.34$                                            & \cite{Freund2004} \\
      \hline
      \multirow{3}{*}{\shortstack{current collector \\ cathode $\Omega_\text{cc}$}} & electronic conductivity  & $\sigma$                               & $3.77 \cdot 10^7 \ \frac{\text{S}}{\text{m}}$     & \cite{JensFreudenberger2020} \\
                                                                 & Young's modulus          & $E$                                    & $7 \cdot 10^{10} \ \text{Pa}$                     & \cite{Freund2004} \\
                                                                 & Poisson's ratio          & $\nu$                                  & $0.34$                                            & \cite{Freund2004} \\
      \hline
      \shortstack{interface current \\ collector - electrode $\Gamma_\text{cs-ed}$}    & interface resistance     & $r_\text{i}$                                    & $2 \cdot 10^{-3} \ \Omega \text{m}^2$                       & defined \\
      \hline
      \shortstack{interface cathode - \\ coating $\Gamma_\text{coat-c}$}                & exchange current density & $i_0$                                  & $4.98 \ \frac{\text{A}}{\text{m}^2}$              & \cite{Schmidt2022} and \cite{Neumann2020} \\
      \hline
      \shortstack{interface anode - \\ solid electrolyte $\Gamma_\text{e-el}$}        & exchange current density & $i_0$                                  & $8.87 \ \frac{\text{A}}{\text{m}^2}$              & \cite{Neumann2020} \\
      \hline
      \shortstack{interface coating - \\ solid electrolyte $\Gamma_\text{el-coat}$}      & interface resistance     & $r_\text{i}$                                    & $5.0 \cdot 10^{-3} \ \Omega \text{m}^2$                    & range reported in~\cite{Javed2020} \\
      \hline
    \end{tabular}
    \label{table:material_parameters_realistic}
\end{table}

%% file: figures/OCP_NMC622.tikz
\definecolor{mycolor1}{rgb}{0.00000,0.44700,0.74100}%

% generated from: Kremer_2019_OCP_NMC622_shifted_new.csv
\begin{tikzpicture}

\begin{axis}[%
width=7.5cm,
height=5cm,
scale only axis,
xmin=0.3,
xmax=1.0,
x tick label style={
/pgf/number format/.cd,
fixed,
fixed zerofill,
precision=1,
/tikz/.cd,
yshift=-.5em},
xlabel style={font=\color{white!15!black}},
xlabel={Lithiation state $\chi$},
ymin=2.5,
ymax=4.5,
ytick={2.5,3.0,3.5,4.0,4.5},
scaled y ticks = false,
y tick label style={
/pgf/number format/.cd,
fixed,
fixed zerofill,
precision=1,
/tikz/.cd},
ylabel style={font=\color{white!15!black}},
ylabel={Open circuit potential $\Phi_0$ in V},
axis background/.style={fill=white},
xmajorgrids,
ymajorgrids,
legend style={at={(0.97,0.03)}, anchor=south east, legend cell align=left, align=left, draw=white!15!black}
]
\addplot [color=mycolor1, line width=1.0pt]
  table[row sep=crcr]{%
0.3 4.47619927240105\\
0.307070707070707 4.45582451376056\\
0.314141414141414 4.43574107056826\\
0.321212121212121 4.41594886951643\\
0.328282828282828 4.39644765830068\\
0.335353535353535 4.37723702746023\\
0.342424242424242 4.35831642956736\\
0.349494949494949 4.33968519611955\\
0.356565656565657 4.32134255243552\\
0.363636363636364 4.30328763081285\\
0.370707070707071 4.28551948216771\\
0.377777777777778 4.26803708634665\\
0.384848484848485 4.25083936127388\\
0.391919191919192 4.23392517107541\\
0.398989898989899 4.21729333330258\\
0.406060606060606 4.20094262536124\\
0.413131313131313 4.18487179023908\\
0.42020202020202 4.16907954161204\\
0.427272727272727 4.15356456840006\\
0.434343434343434 4.13832553883432\\
0.441414141414141 4.12336110408984\\
0.448484848484848 4.10866990153148\\
0.455555555555555 4.09425055761504\\
0.462626262626263 4.08010169048072\\
0.46969696969697 4.06622191227157\\
0.476767676767677 4.05260983120601\\
0.483838383838384 4.03926405342996\\
0.490909090909091 4.02618318467165\\
0.497979797979798 4.0133658317191\\
0.505050505050505 4.00081060373859\\
0.512121212121212 3.98851611345006\\
0.519191919191919 3.97648097817383\\
0.526262626262626 3.96470382076164\\
0.533333333333333 3.95318327042322\\
0.54040404040404 3.94191796345897\\
0.547474747474747 3.93090654390784\\
0.554545454545454 3.92014766411865\\
0.561616161616162 3.90963998525251\\
0.568686868686869 3.89938217772274\\
0.575757575757576 3.88937292157861\\
0.582828282828283 3.87961090683814\\
0.58989898989899 3.87009483377496\\
0.596969696969697 3.86082341316362\\
0.604040404040404 3.85179536648736\\
0.611111111111111 3.84300942611191\\
0.618181818181818 3.8344643354287\\
0.625252525252525 3.82615884897028\\
0.632323232323232 3.81809173250078\\
0.639393939393939 3.81026176308373\\
0.646464646464646 3.80266772912953\\
0.653535353535354 3.79530843042446\\
0.660606060606061 3.78818267814323\\
0.667676767676768 3.78128929484642\\
0.674747474747475 3.77462711446467\\
0.681818181818182 3.76819498227066\\
0.688888888888889 3.76199175484031\\
0.695959595959596 3.75601630000432\\
0.703030303030303 3.75026749679087\\
0.71010101010101 3.74474423536075\\
0.717171717171717 3.73944541693546\\
0.724242424242424 3.7343699537192\\
0.731313131313131 3.72951676881543\\
0.738383838383838 3.72488479613869\\
0.745454545454545 3.72047298032212\\
0.752525252525252 3.71628027662136\\
0.75959595959596 3.71230565081522\\
0.766666666666667 3.70854807910363\\
0.773737373737374 3.70500654800316\\
0.780808080808081 3.70168005424053\\
0.787878787878788 3.6985676046444\\
0.794949494949495 3.69566821603546\\
0.802020202020202 3.69298091511488\\
0.809090909090909 3.69050473835049\\
0.816161616161616 3.68823873185844\\
0.823232323232323 3.68618195127564\\
0.83030303030303 3.6843334616095\\
0.837373737373737 3.68269233703302\\
0.844444444444444 3.68125766054535\\
0.851515151515151 3.68002852330237\\
0.858585858585859 3.67900402313585\\
0.865656565656566 3.6781832610791\\
0.872727272727273 3.67756533299432\\
0.87979797979798 3.67714930916538\\
0.886868686868687 3.67693418432381\\
0.893939393939394 3.67691875503528\\
0.901010101010101 3.6771013186283\\
0.908080808080808 3.67747893369498\\
0.915151515151515 3.67804560348254\\
0.922222222222222 3.67878781309469\\
0.929292929292929 3.67967356566211\\
0.936363636363636 3.68062544709972\\
0.943434343434343 3.68145445308287\\
0.95050505050505 3.68169741857838\\
0.957575757575758 3.68021762288952\\
0.964646464646465 3.67422357599325\\
0.971717171717172 3.65685842139712\\
0.978787878787879 3.61127770022431\\
0.985858585858586 3.49609989452052\\
0.992929292929293 3.20966204045106\\
1 2.50220468352966\\
};

\end{axis}
\end{tikzpicture}%